\documentclass[twocolumn]{aastex62}
\usepackage{import}

\usepackage{mathtools}
\usepackage{chngcntr}
\usepackage{bm}
\usepackage{amsmath}
\usepackage{amssymb}
\usepackage{chngcntr}
\usepackage{footnote}
\usepackage[caption=false]{subfig}
\usepackage{float}
\usepackage{tabulary,booktabs}
\usepackage{multirow}
\usepackage{yhmath}
\usepackage{graphicx}
\usepackage{csquotes}
\usepackage{mwe}

\shorttitle{Clumps in UVCANDELS}
\shortauthors{Sattari et al.}
\begin{document}
\title{\large \textbf{Fraction of Clumpy Star-Forming Galaxies at $0.5\leq z\leq 3$ in UVCANDELS: Dependence on Stellar Mass and Environment}}

\correspondingauthor{Zahra Sattari}
\email{zahra.sattari@email.ucr.edu}\\
\email{zsattari@carnegiescience.edu}

\author[0000-0002-0364-1159]{Zahra Sattari}
\affiliation{Department of Physics and Astronomy, University of California, Riverside, 900 University Ave, Riverside, CA 92521, USA}
\affiliation{The Observatories of the Carnegie Institution for Science, 813 Santa Barbara St., Pasadena, CA 91101, USA}

\author{Bahram Mobasher}
\affiliation{Department of Physics and Astronomy, University of California, Riverside, 900 University Ave, Riverside, CA 92521, USA}

\author[0000-0003-3691-937X]{Nima Chartab}
\affiliation{The Observatories of the Carnegie Institution for Science, 813 Santa Barbara St., Pasadena, CA 91101, USA}

\author[0000-0003-4727-4327]{Daniel D. Kelson}
\affiliation{The Observatories of the Carnegie Institution for Science, 813 Santa Barbara St., Pasadena, CA 91101, USA}

\author[0000-0002-7064-5424]{Harry I. Teplitz}
\affiliation{IPAC, Mail Code 314-6, California Institute of Technology, 1200 E. California Blvd., Pasadena, CA, 91125, USA}

\author[0000-0002-9946-4731]{Marc Rafelski}
\affiliation{Space Telescope Science Institute, 3700 San Martin Drive, Baltimore, MD 21218, USA}
\affiliation{Department of Physics and Astronomy, Johns Hopkins University, Baltimore, MD 21218, USA}

\author[0000-0001-9440-8872]{Norman A. Grogin}
\affiliation{Space Telescope Science Institute, 3700 San Martin Drive, Baltimore, MD 21218, USA}

\author[0000-0002-6610-2048]{Anton M. Koekemoer}
\affiliation{Space Telescope Science Institute, 3700 San Martin Drive, Baltimore, MD 21218, USA}

\author[0000-0002-9373-3865]{Xin Wang}
\affiliation{School of Astronomy and Space Science, University of Chinese Academy of Sciences (UCAS), Beijing 100049, China}
\affiliation{National Astronomical Observatories, Chinese Academy of Sciences, Beijing 100101, China}

\author[0000-0001-8156-6281]{Rogier A. Windhorst}
\affiliation{Arizona State University, School of Earth \& Space Exploration, Tempe, AZ 85287-1404, USA}

\author[0000-0002-8630-6435]{Anahita Alavi}
\affiliation{IPAC, Mail Code 314-6, California Institute of Technology, 1200 E. California Blvd., Pasadena, CA, 91125, USA}

\author[0000-0002-0604-654X]{Laura Prichard}
\affiliation{Space Telescope Science Institute, 3700 San Martin Drive, Baltimore, MD 21218, USA}

\author[0000-0003-3759-8707]{Ben Sunnquist}
\affiliation{Space Telescope Science Institute, 3700 San Martin Drive, Baltimore, MD 21218, USA}

\author[0000-0003-2098-9568]{Jonathan P. Gardner}
\affiliation{Astrophysics Science Division, NASA Goddard Space Flight Center, 8800 Greenbelt Rd, Greenbelt, MD 20771, USA}

\author[0000-0003-1530-8713]{Eric Gawiser}
\affiliation{Department of Physics and Astronomy, Rutgers, the State University of New Jersey, Piscataway, NJ 08854, USA}

\author[0000-0001-6145-5090]{Nimish P. Hathi}
\affiliation{Space Telescope Science Institute, 3700 San Martin Drive, Baltimore, MD 21218, USA}

\author[0000-0001-8587-218X]{Matthew J. Hayes}
\affiliation{Department of Astronomy and Oskar Klein Centre, Stockholm University, AlbaNova University Centre, SE-10691, Stockholm, Sweden}

\author[0000-0001-7673-2257]{Zhiyuan Ji}
\affiliation{Steward Observatory, University of Arizona, 933 N. Cherry Ave, Tucson, AZ 85719, USA}

\author[0000-0001-7166-6035]{Vihang Mehta}
\affiliation{IPAC, Mail Code 314-6, California Institute of Technology, 1200 E. California Blvd., Pasadena, CA, 91125, USA}

\author[0000-0002-4271-0364]{Brant E. Robertson}
\affiliation{Department of Astronomy and Astrophysics, University of California, Santa Cruz, 1156 High Street, Santa Cruz, CA 95064, USA}

\author[0000-0002-9136-8876]{Claudia Scarlata}
\affiliation{Minnesota Institute of Astrophysics and School of Physics and Astronomy, University of Minnesota, Minneapolis, MN, USA}

\author[0000-0003-3466-035X]{L. Y. Aaron\ Yung}
\affiliation{Astrophysics Science Division, NASA Goddard Space Flight Center, 8800 Greenbelt Rd, Greenbelt, MD 20771, USA}

\author[0000-0003-1949-7638]{Christopher J. Conselice}
\affiliation{Jodrell Bank Centre for Astrophysics, University of Manchester, Oxford Road, Manchester, UK}

\author[0000-0002-7928-416X]{Y. Sophia Dai}
\affiliation{Chinese Academy of Sciences South America Center for Astronomy (CASSACA), National Astronomical Observatories(NAOC),
20A Datun Road, Beijing 100012, China}

\author[0000-0003-2775-2002]{Yicheng Guo}
\affiliation{Department of Physics and Astronomy, University of Missouri, Columbia, MO 65211, USA}

\author[0000-0003-1581-7825]{Ray A. Lucas}
\affiliation{Space Telescope Science Institute, 3700 San Martin Drive, Baltimore, MD 21218, USA}

\author[0000-0002-6632-4046]{Alec Martin}
\affiliation{Department of Physics and Astronomy, University of Missouri, Columbia, MO 65211, USA}

\author[0000-0002-5269-6527]{Swara Ravindranath}
\affiliation{Space Telescope Science Institute, 3700 San Martin Drive, Baltimore, MD 21218, USA}

\begin{abstract}

High-resolution imaging of galaxies in rest-frame UV has revealed the existence of giant star-forming clumps prevalent in high redshift galaxies. Studying these sub-structures provides important information about their formation and evolution and informs theoretical galaxy evolution models. We present a new method to identify clumps in galaxies' high-resolution rest-frame UV images. Using imaging data from CANDELS and UVCANDELS, we identify star-forming clumps in an HST/F160W$\leq 25$ AB mag sample of 6767 galaxies at $0.5\leq z\leq 3$ in four fields, GOODS-N, GOODS-S, EGS, and COSMOS. We use a low-pass band filter in Fourier space to reconstruct the background image of a galaxy and detect small-scale features (clumps) on the background-subtracted image. Clumpy galaxies are defined as those having at least one off-center clump that contributes a minimum of 10\% of the galaxy's total rest-frame UV flux. We measure the fraction of clumpy galaxies ($\rm f_{clumpy}$) as a function of stellar mass, redshift, and galaxy environment. Our results indicate that $\rm f_{clumpy}$ increases with redshift, reaching $\sim 65\%$ at $z\sim 1.5$. We also find that $\rm f_{clumpy}$  in low-mass galaxies ($\rm 9.5\leq log(M_*/M_\odot)\leq 10$) is 10\% higher compared to that of their high-mass counterparts ($\rm log(M_*/M_\odot)>10.5$). Moreover, we find no evidence of significant environmental dependence of $\rm f_{clumpy}$ for galaxies at the redshift range of this study. Our results suggest that the fragmentation of gas clouds under violent disk instability remains the primary driving mechanism for clump formation, and incidents common in dense environments, such as mergers, are not the dominant processes.

\end{abstract}

\section{Introduction}

High-redshift star-forming galaxies (SFGs) frequently harbor compact regions of star formation, known as ``clumps'', which are abundant in cool gas \citep[e.g.,][]{Conselice04,Elmegreen05,Elmegreen07,Bournaud08,Schreiber11}. These regions are fed by cold gas inflow from intergalactic medium (IGM) into the galaxy \citep{Dekel06,Dekel09n}, and their co-evolution with their host galaxy is affected by various processes within the galaxy or by incidents in the local environment \citep[e.g.,][]{Dekel09,Mandelker14,Mandelker17}. Several observational studies and simulations have suggested two commonly accepted scenarios for the formation of these clouds: either they are formed through the fragmentation of gas clouds under gravitational instability of the violent disk (\textit{in-situ}), or gas-rich minor mergers (\textit{ex-situ}). The formation of clumps via violent disk instability (VDI) is supported by different simulations \citep[e.g.,][]{Noguchi99,Immeli04a,Immeli04b,Bournaud07,Bournaud09,Elmegreen08,Genzel08,Genzel11,Dekel09,Agertz09,Ceverino10,Ceverino12,Inoue16} and observations \citep[e.g.,][]{Elmegreen07,Guo12,Guo15,Shibuya16,Hinojosa16,Mieda16,Soto17,Fisher17,Zanella19,Adams22,Sok22}. However, there are also studies that indicate mergers as the origin of clump formation \citep[e.g.,][]{Robertson08,Puech10,Hopkins13,Straughn15}.

Clumps are reported as areas with elevated specific star formation rates (sSFR; defined as star formation rate (SFR) per stellar mass) compared to their surrounding regions \citep{Guo12,Wuyts12,Hemmati14,Mieda16,Mehta21,Iani21}, with majority of them in the mass range $\rm 10^7-10^9\ M_\odot$ \citep{Elmegreen07,Guo12,Guo15,Soto17,Zavadsky17mass,Huertas20,Ambachew22}. However, studies of high-redshift lensed systems revealed smaller stellar masses ($\rm 10^6\ M_\odot$) and sizes ($\leq 100$ pc) for clumps \citep{Livermore15,Johnson17,Rigby17,Cava18,Zavadsky18,Zavadsky19,Zick20,Vanzella21,Vanzella22,mestric22,Messa22,Claeyssens23} since, by utilizing these systems, we are able to achieve a significant increase in resolution, which allows for the investigation of much smaller scales than what is possible with field galaxies \citep{Livermore12,Adamo13,Johnson17,Vanzella19,Sharma21,mestric22,Welch23}.

Rest-frame ultraviolet (UV) images of galaxies trace their SFR over timescale of $\sim 100$ Myr associated with continuum from massive, short-lived  ${\rm{O}}$- and ${\rm{B}}$-type stars \citep{Calzetti13}. Thus, clumps can be detected in high-resolution imaging of field or lensed galaxies in rest-frame UV or optical \citep[e.g.,][]{Conselice04,Elmegreen05,Elmegreen07,Taylor-Mager07,Elmegreen09,Schreiber11,Guo12,Murata14,Tadaki14,Guo15,Shibuya16,Soto17,Cava18,Mager18,Guo18,Zavadsky18,Messa19,Vanzella21,Sok22,mestric22}. There are, however, other studies that have identified clumps in $\rm H\alpha$ emission line maps of galaxies \citep[e.g.,][]{Genzel08,Genzel11,Livermore12,Mieda16,Fisher17,Zanella19,Sharma21} or in CO observations of lensed galaxies \citep[e.g.,][]{Jones10,Swinbank10}. In this work, we define clumps as off-center star-forming regions that can be detected in the rest-frame UV images of galaxies.

Clumps can also evolve within galaxies and contribute to their morphology. The evolution of the clumps within their host galaxies has been the subject of debate. Different observations and simulations support various scenarios, such as the migration of clumps towards the center of the galaxy and forming the progenitor of the galaxy's present bulge due to dynamical friction or clump-clump interaction \citep[e.g.,][]{Dekel09,Ceverino12,Mandelker14,Shibuya16,Soto17,Mandelker17,Mehta21,Dekel22}. However, other simulations found that stellar feedback can disturb these clumps and even destroy them before migrating to the center and thus, they have a short lifetime. In this scenario, disrupted clumps contribute to the formation of thick disks in their host galaxies \citep[e.g.,][]{Murray10,Genel12,Hopkins12,Moody14}.

Despite our current understanding, the formation and evolution of clumps in galaxies remains poorly understood and requires further study to help refine and constrain theoretical models. By examining clumps in galaxies, we can gain valuable insights into the history of their host galaxies. Moreover, their study has the benefit of testing the validity of feedback models in simulations. However, observations of clumps are challenging due to the limited spatial resolution specially of high redshift galaxies. With the advent of sensitive detectors on space telescopes, we can obtain multi-waveband imaging data with much higher spatial resolution for high-redshift galaxies.

In this paper, we identify star-forming clumpy regions in the rest-frame 1600 \r{A} images of galaxies using data from The Cosmic Assembly Near-IR Deep Extragalactic Legacy Survey \citep[CANDELS;][]{Grogin11,Koekemoer11} and Ultraviolet Imaging of the Cosmic Assembly Near-infrared Deep Extragalactic Legacy Survey (UVCANDELS; Wang et al. in preparation) in the redshift range of $0.5 \leq z \leq 3$. We introduce a new method to subtract the smooth component of galaxy images and detect clumps in the residual images. We then investigate how the fraction of clumpy galaxies changes as a function of galaxy properties, such as stellar mass and environment. Studying such scaling relations can inform us about the details of clump formation within galaxies. Mainly, the environmental dependence of the clumpy fraction is studied for the first time and can test scenarios regarding formation history of clumps via mergers. Moreover, we study the redshift evolution of clumpy galaxies and compare our results with previous studies. We discuss the evolutionary path of clumpy galaxies and address discrepancies between our work and previous studies.

The paper is structured as follows. In Section \ref{Data}, we describe the data and the selection of our sample. We present clump identification method in Section \ref{Clump_identification} and examine completeness of our technique. In Section \ref{Results} we investigate the measurements of clumpy fraction and their evolution with different global properties of galaxies. The fraction of clumpy galaxies in relation to redshift is also investigated in this section and we compare our results with other studies in the literature. We discuss the physical interpretation of our results and summarize them in Section \ref{Discussion}.

Throughout this paper, we assume a flat $\Lambda$CDM cosmology with $H_0=70\ \rm kms^{-1} Mpc^{-1}$, $\Omega_{m_{0}}=0.3$ and $\Omega_{\Lambda_{0}}=0.7$. All the physical parameters are measured assuming a \cite{Chabrier03} initial mass function (IMF) and magnitudes are represented in the AB system.
\section{Data}\label{Data}

\subsection{Catalogs and Mosaics}

We analyze rest-frame UV images of galaxies to identify star-forming clumps. We select galaxies from multi-wavelength catalogs in four CANDELS fields: GOODS-S \citep{Guo13}, GOODS-N \citep{Barro19}, COSMOS \citep{Nayyeri17}, and EGS \citep{Stefanon17}. Each of GOODS-S and GOODS-N fields covers an area of $170\ \rm arcmin^2$ with $5 \sigma$ limiting AB magnitude depth of 27.36 (GOODS-S) and 27.8 (GOODS-N) in F160W band. Also, the areal coverage of COSMOS and EGS fields is 216 and 206 $\rm arcmin^2$ with $5 \sigma$ limiting AB magnitude depth (in F160W band) of 27.56 and 27.6, respectively.

\begin{figure}
\centering
\includegraphics[width=0.48\textwidth,clip=True, trim=0cm 0cm 0cm 0cm]{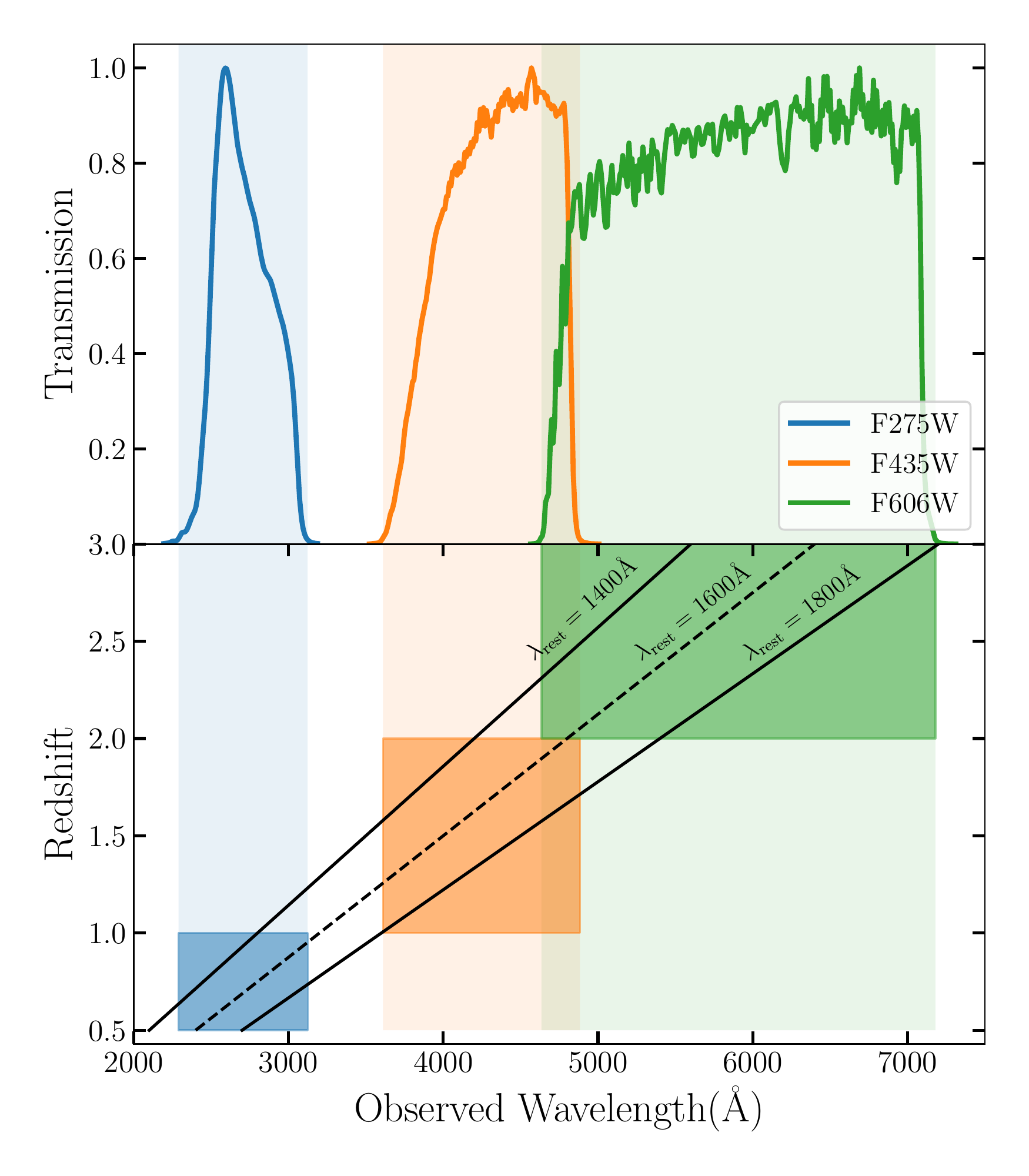}
\caption{\textit{Top}: Filter throughput (normalized to the peak transmission) for three wavebands that are utilized in this work. The shaded regions in the figure show the filter width with at least 1\% of maximum transmission. \textit{Bottom}: The observed wavelength of rest-frame 1400, 1600 and 1800 \r{A} lines as a function of redshift. Blue, orange and green boxes correspond to rest-frame UV coverage for galaxies at the redshift range of $0.5-1, 1-2$, and $2-3$, respectively.}\label{throughput} 
\end{figure}

Recently, UVCANDELS survey with the Hubble Space Telescope (PI: H. Teplitz, Cycle 26 GO 15647) conducted UV observations of these four fields in WFC3/UVIS F275W and ACS/WFC F435W bands. For the redshift range of this study ($0.5 \leq z \leq 3$), F275W, F435W, and F606W bands probe the rest-frame UV 1600 \r{A} at $0.5 \leq z <1$, $1 \leq z<2$, and $2 \leq z \leq 3$, respectively (Figure \ref{throughput}). We use 60-mas mosaics of aforementioned bands to identify clumps. Although 30-mas mosaics can also be used due to the smaller pixel scale of the ACS instrument, 60-mas images provide sufficient resolution to probe our desired clumps with the size of $\sim$kpc, which contribute significantly to the total SFR of galaxies (Section \ref{Clump_identification}). The CANDELS/UDS field is not included in this study since it does not have observations in two of these bands (F275W and F435W).

To calculate the physical properties of galaxies such as their stellar mass ($\rm M_*$) and SFR, the spectral energy distribution (SED) fitting is performed using Code Investigating GALaxy Emission \citep[CIGALE;][]{Boquien19}. In the SED fitting procedure, the new observations of F275W and F435W bands obtained by UVCANDELS are also added to the existing data. Moreover, photometric redshifts of the galaxies are taken from the UVCANDELS catalog that includes F275W and F435W photometry. The catalog combines the redshift probability distributions of three different codes to compute robust values for the photometric redshifts (Sunnquist et al. in preparation).

A detailed description of the SED fitting procedure will be published in a future paper (Mehta et al. in preparation). In brief, the synthetic spectral library of \cite{Bruzual03} with the following assumptions is used to perform the SED fitting: A delayed exponentially declining star formation history ($t e^{-t/\tau}$) with a range of 30 Myr to 30 Gyr e-folding time-scales ($\tau$) is considered. Additionally, we allow for the possibility of an episode of recent star-burst as a 10 Myr old burst with an exponential e-folding time of 50 Myr and the contribution of the burst is parameterized by the fraction of total mass generated in the burst. The code also includes the contribution of nebular emission lines. The stellar metallicities of Z = 0.0001, 0.0004, 0.004, 0.008, 0.02, 0.05, and a \cite{Calzetti00} extinction law are adopted to generate the template SEDs and perform the fitting.


\subsection{Sample Selection}\label{sample_selection}

The sample of galaxies we study in this paper consists of HST/F160W (H band)-selected sources, in four CANDELS fields: GOODS-S, GOODS-N, COSMOS, and EGS. Target galaxies are all covered by the desired HST/ACS and HST/WFC3 imaging data, i.e., galaxies with $0.5\leq z < 1$, $1\leq z < 2$ and $2\leq z \leq3$ have coverage in F275W, F435W and F606W bands, respectively.

SFRs derived from SED fitting are well-constrained when rest-frame UV data are available. We, therefore, utilize a threshold defined by \cite{Pacifici16} on the sSFR of galaxies to select a sample of star-forming galaxies. Based on this threshold, galaxies with sSFR$>0.2/t_U(z)$ are identified as star-forming galaxies, where $t_U(z)$ is the age of the Universe at redshift $z$. We investigated the alternative UVJ color-color selection and obtained a similar sample. Moreover, we require all galaxies to have $\rm F160W \leq 25$ mag and $\rm M_*\geq 10^{9.5}M_\odot$ over the redshift range of $0.5\leq z \leq 3$. The stellar mass limit is imposed to obtain a mass-complete sample within the redshift range of this study. 

\begin{table}
\centering

\caption{Summary of data utilized in this study.}

\begin{tabular}{@{\extracolsep{\fill}}cccc@{}}

\hline
Field & Area ($\rm arcmin^2$) & $5 \sigma$ Depth (AB) & N \footnote{Number of galaxies in each field in the final sample.} \\ \hline \hline
GOODS-S       & 170                            & 27.36                 & 1572                                    \\
GOODS-N        & 170                                    & 27.8                & 1870                             \\
COSMOS        & 216                                    & 27.56                & 1304                             \\
EGS       & 206                                 & 27.6   & 2021 \\ \hline
\end{tabular}%
\label{tab1}
\end{table}

Active galactic nuclei (AGNs) as well as stars (stellarity parameter of {\tt SExtractor} \citep{Bertin96}, $\rm CLASS\_STAR >0.9$) are excluded from our sample. Also, to allow measurements of resolved images and identification of clumps, we select face-on galaxies with axial ratio (q) $>0.5$ in their F160W images (the ratio of the major and minor axes). Furthermore, we require the size of the semi-major axis (SMA) of the galaxies to be $>0.2 {\arcsec}$. All the above criteria lead us to a sample of 1572, 1870, 2021, and 1304 galaxies in the GOODS-S and GOODS-N, EGS, and COSMOS fields, respectively. Table \ref{tab1} summarizes detailed information about the data and sample size.

\begin{figure*}
    \centering
    \subfloat{\includegraphics[width=0.95\textwidth,clip=True, trim=0cm 0cm 0cm 0cm]{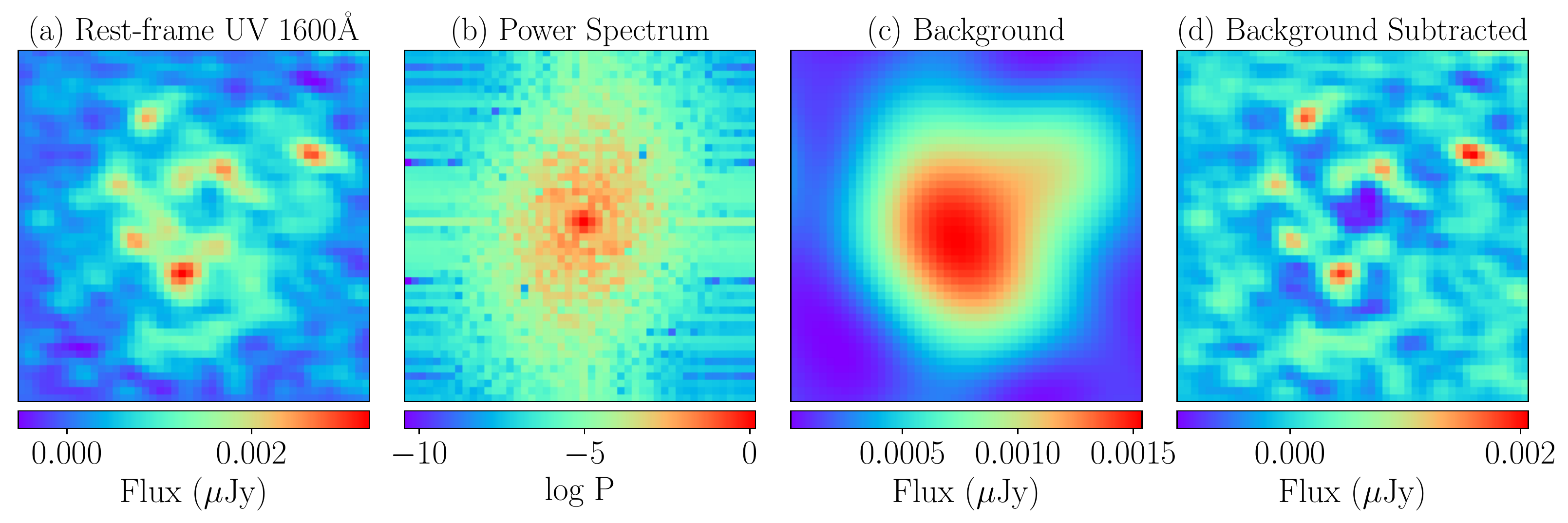}}
    \qquad
    \subfloat{\includegraphics[width=0.95\textwidth,clip=True, trim=0cm 0cm 0cm 0cm]{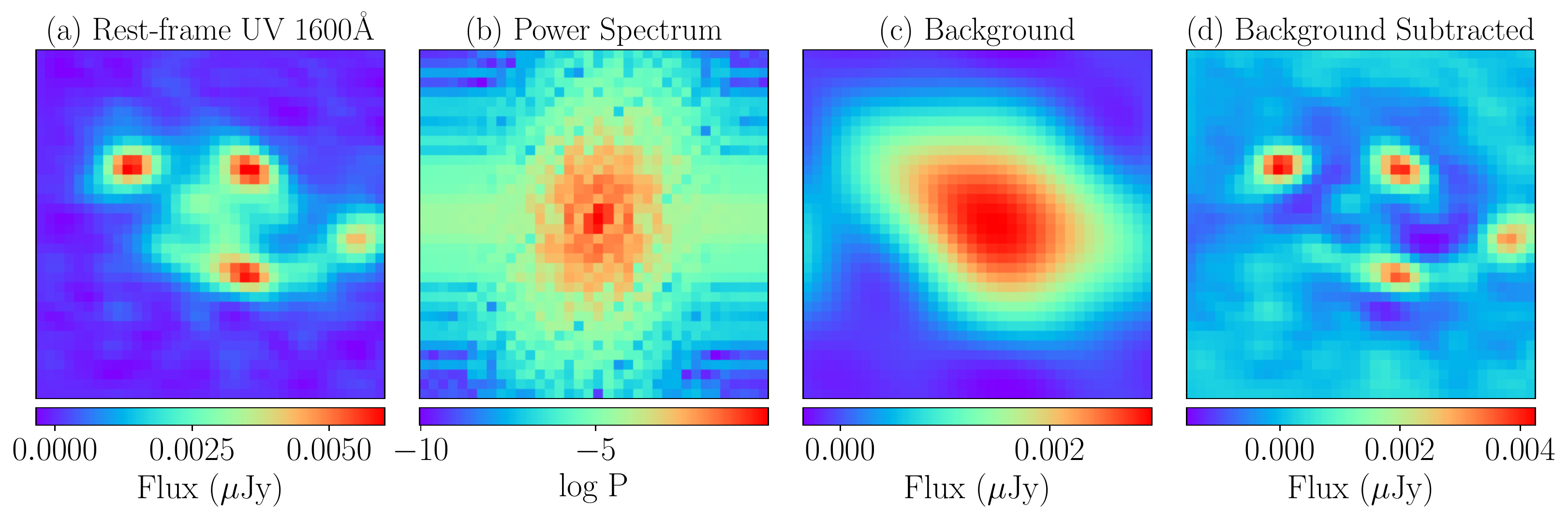}}
    \caption{Two examples demonstrating the process of subtracting background from rest-frame UV images of galaxies. Panel (a) shows the galaxy image in the rest-frame UV filter. We calculate the power spectrum of this image in the Fourier space. Panel (b) shows log(power spectrum) in the frequency domain. After constructing the background map of the clump (Panel (c)), we remove it from the original image and the residual is an image which is ready to identify its clumps (Panel (d)).}\label{power_spectrum} 
\end{figure*}

\section{Clump Identification}\label{Clump_identification}

We detect clumpy regions in the rest-frame UV 1600 \r{A} images of galaxies. The sample is divided into three redshift bins and we inspect images of galaxies in a filter that corresponds to the rest-frame UV wavelength in each redshift bin. Thus, we detect the clumps in F275W, F435W and F606W bands at the redshift ranges $0.5\leq z <1$, $1\leq z <2$, and $2\leq z\leq 3$, respectively. We PSF-match all mosaics to the F160W band to achieve three goals. Firstly, to accurately distinguish clumps from bulges of host galaxies, it is beneficial to have a consistent definition of their centers, which are defined in the F160W band. Secondly, to minimize noise and ensure consistent clump detection throughout the redshift range, matching the PSF of the other filters with that of the F160W band is advantageous. The FWHM of the F160W PSF is 0.17 arcseconds ($<3$ pixels), which is less than the minimum number of pixels required to define a clump. Lastly, detection of clumps in PSF-matched images are necessary for future work that involves measuring SED of individual clumps.

\subsection{Method}

We first construct and subtract the local background of clumps in galaxy images. There are various ways to define and remove the background which involve subtracting the smooth background of the image that corresponds to large-scale variations compared to typical clump size. In this work, we use Fourier transformation to decompose small-scale features (i.e., clumps) from the smooth image of a galaxy.

One approach to construct the background of the clumps is to smooth the galaxy image with a custom kernel function (e.g., Gaussian function). This method usually has an underlying assumption about the symmetrical shape of galaxies and involves a hyper-parameter (i.e., the width of a kernel function) which controls the background construction and hence the clump identification. However, the diverse shape of galaxies, especially at high redshifts, needs to be modeled non-parametrically and asymmetrically to develop an effective clump identification technique.  

In order to make an adaptive method to take into account the irregular morphology of the galaxies properly while constructing the background, we use a method based on Fast Fourier Transform (FFT) \citep[SciPy;][]{scipy}. The steps involved in background subtraction are demonstrated in Figure \ref{power_spectrum}. We first transfer the rest-frame UV image of the galaxy to Fourier (frequency) space. We then make the power spectrum map of the galaxy in this space (Panel (b)). This map shows the power distribution coming from various scales (large-scale or small frequencies, and small-scale or high frequencies). The struggle to eliminate the background of the clumps now reduces to masking low frequencies in the FFT image since these frequencies represent the large-scale background features.

\begin{figure*}
\centering
\begin{tabular}{cc}
    \includegraphics[width=0.5\linewidth]{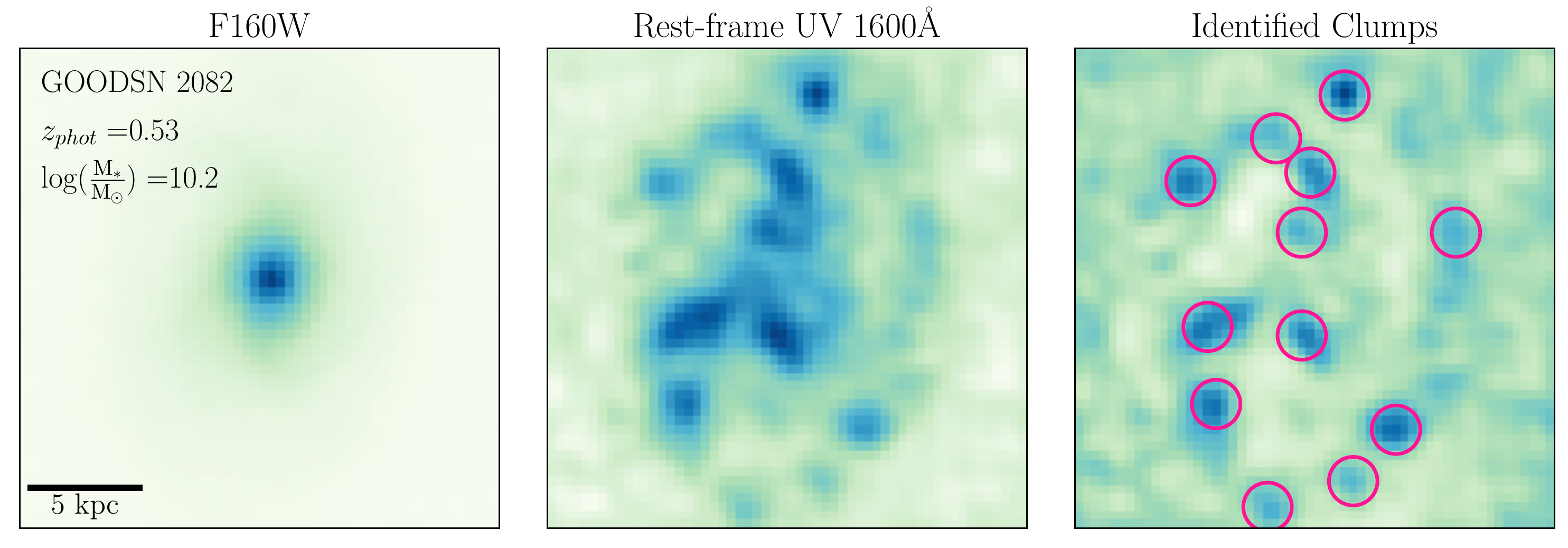}&
    \includegraphics[width=0.5\linewidth]{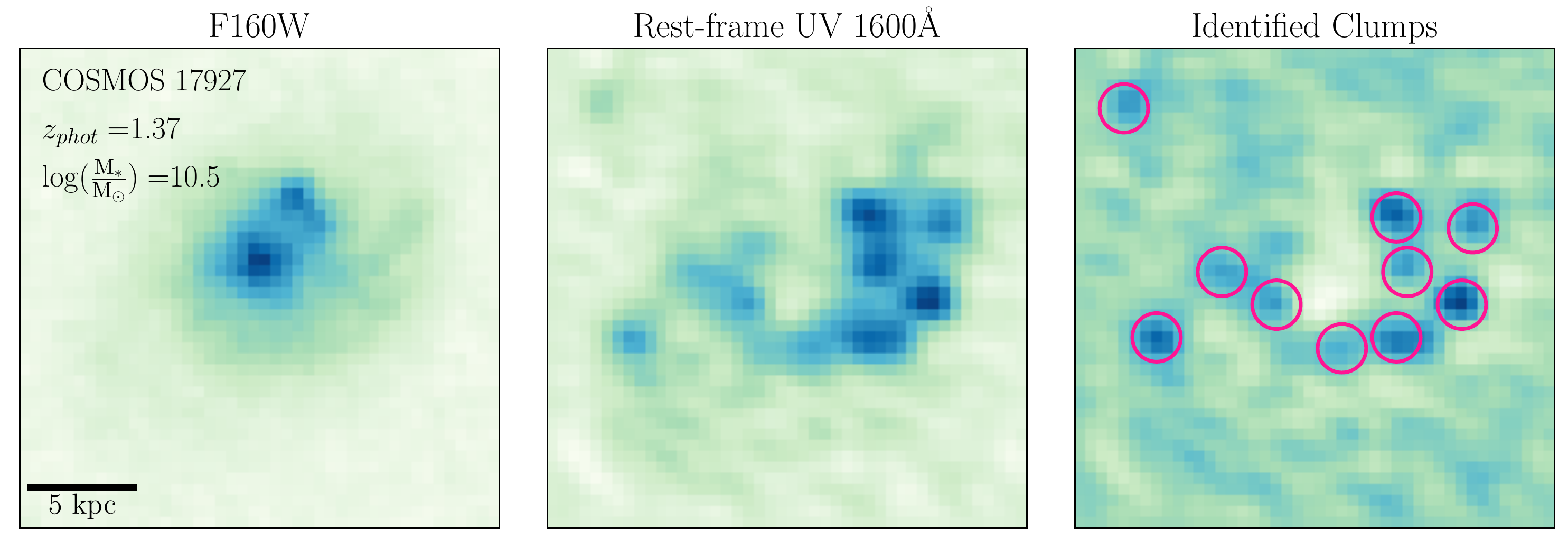}\\[2\tabcolsep]
    \includegraphics[width=0.5\linewidth]{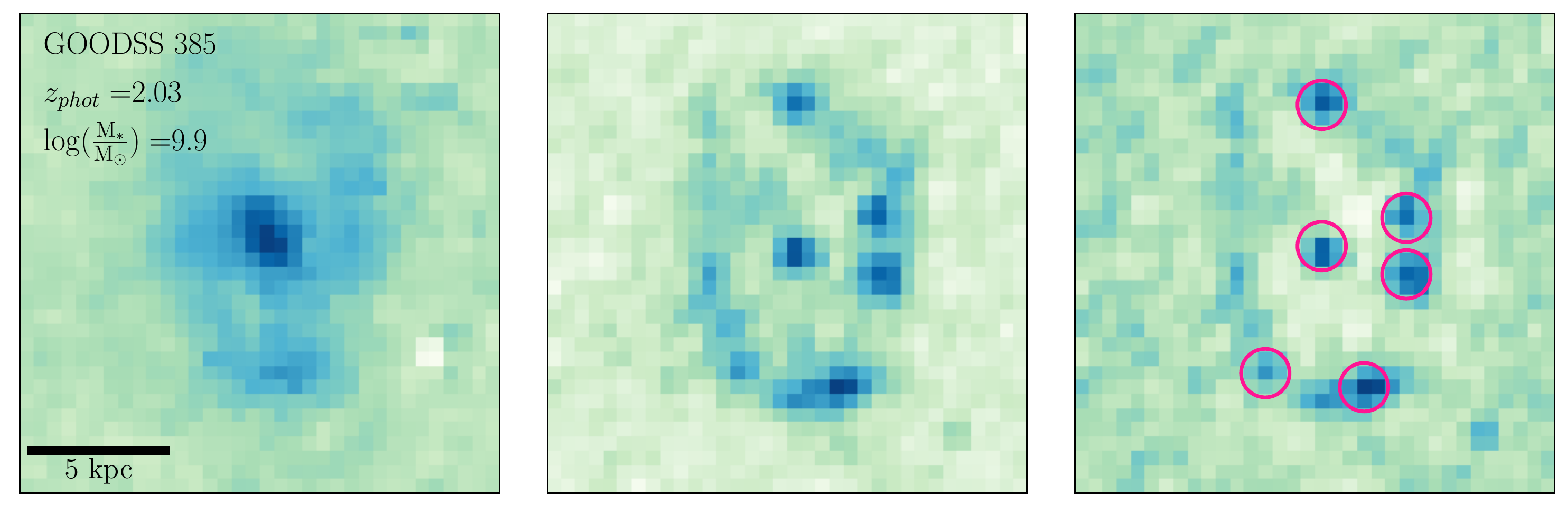}&
    \includegraphics[width=0.5\linewidth]{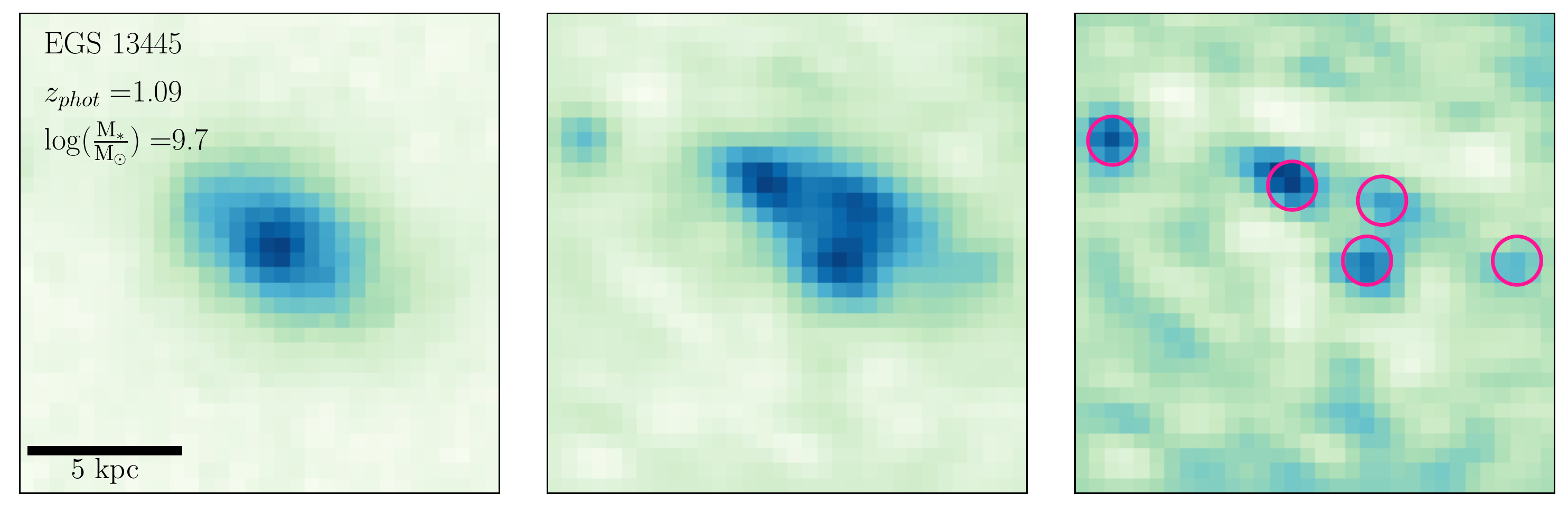}\\[2\tabcolsep]
    \includegraphics[width=0.5\linewidth]{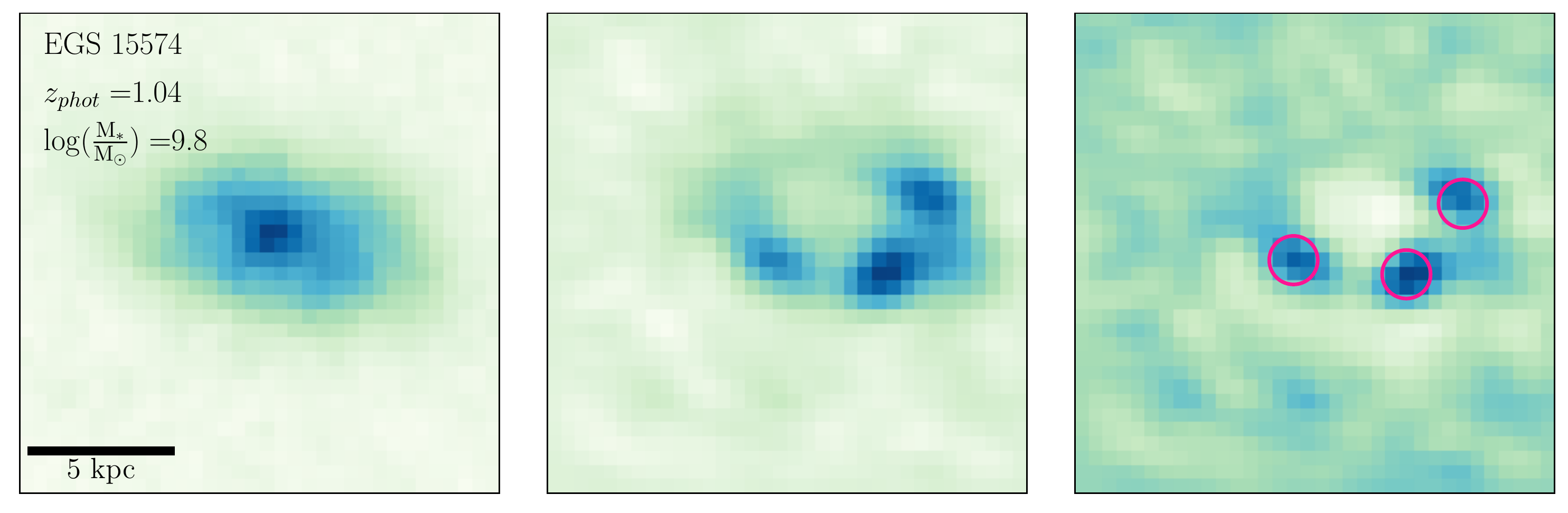}&
    \includegraphics[width=0.5\linewidth]{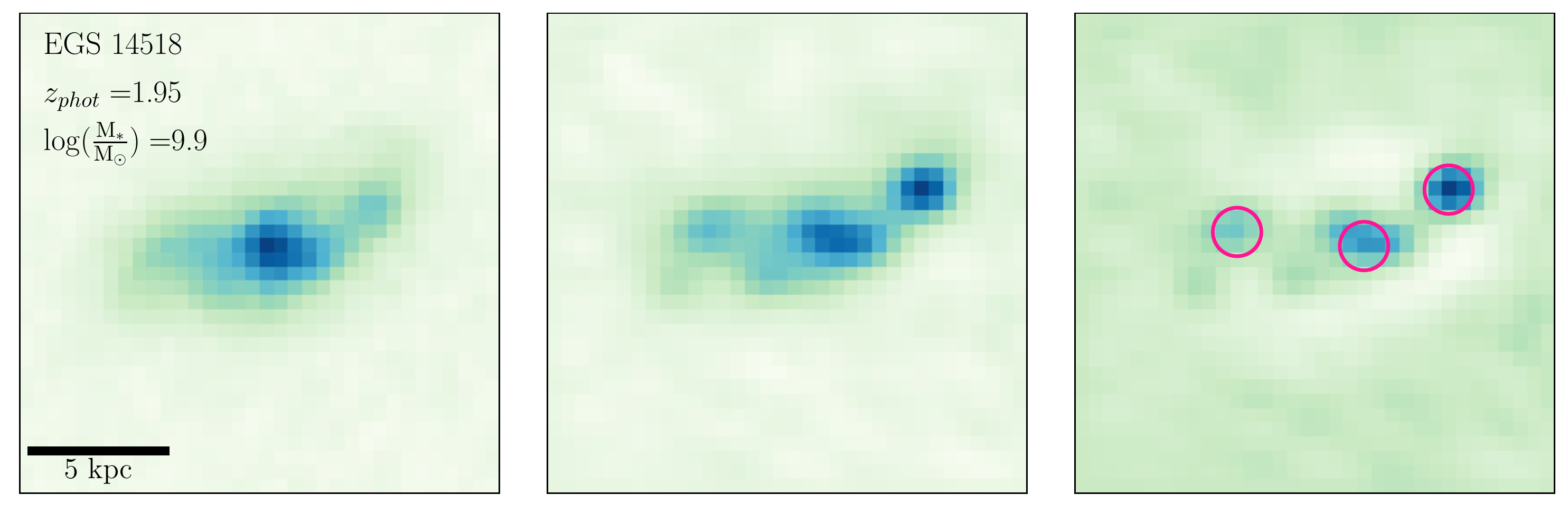}\\[2\tabcolsep]
    \includegraphics[width=0.5\linewidth]{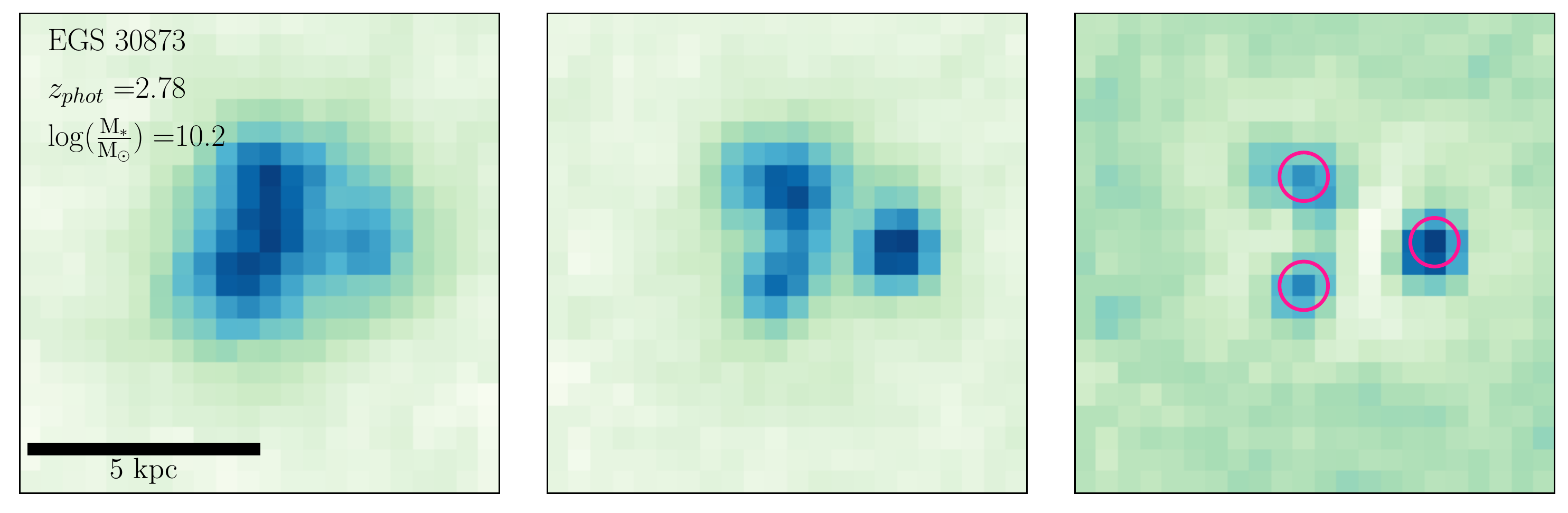}&
    \includegraphics[width=0.5\linewidth]{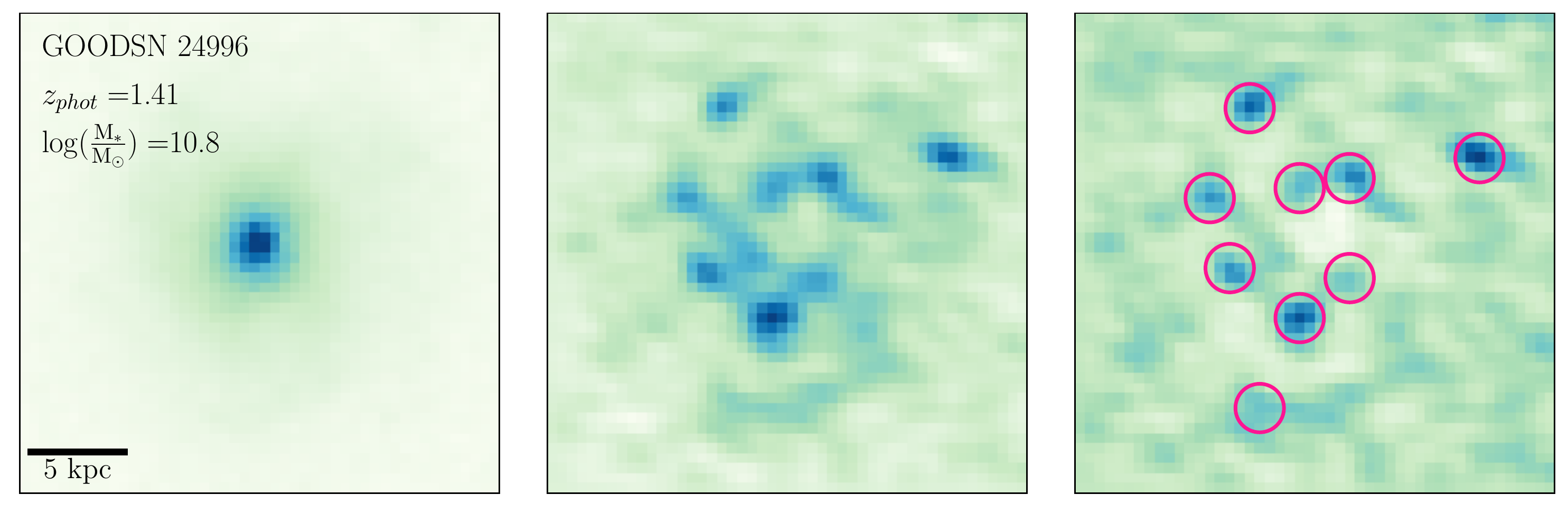}
\end{tabular}
\caption{Eight examples of clumpy galaxies after identifying their clumps with magenta circles on their background-subtracted images in the right panels. Also, the left and middle panels show F160W and rest-frame UV 1600 \r{A} images of these galaxies, respectively. In Section \ref{complete}, we eliminate clumps that account for less than 10\% of the total rest-frame UV flux of their host galaxies, resulting in a complete sample of clumpy galaxies. However, in this figure, we do not apply this requirement.}
\label{example}
\end{figure*}

As previously stated, making a high-pass filter to mask the lower frequencies in the image is challenging because of the variety in the morphology of different galaxies. To effectively remove lower frequencies in an image of a galaxy, we first fit a 2-dimensional asymmetric Gaussian function to the power spectrum image of the galaxy to determine the values of standard deviation in the $x$ and $y$ directions ($\sigma_x$ and $\sigma_y$). Based on these values, we then employ a box high-pass filter (with the size of 2$\sigma_x$ $\times$ 2$\sigma_y$) that is tailored to the specific galaxy. The filter is then convolved with the FFT image of the galaxy to construct the background-subtracted image where we identify clumps (Panel (d) in Figure \ref{power_spectrum}). The background image is shown in panel (c) of Figure \ref{power_spectrum}. Compared to the original image of the galaxy (Panel (a)), we can see most of the features corresponding to large-scale variations are removed in this image.

A similar background subtraction method has been used by \cite{Wang15} who investigated the filamentary structures in Milky Way spiral arms. They considered a threshold of $90\%$ of the maximum of the power spectrum and selected pixels above this cut as representatives of the low spatial frequencies (large-scale features of background). As described above, we avoid a constant threshold and compute it based on the power distribution of a galaxy in the frequency space.

After making a background-subtracted image for each galaxy, we use an image segmentation method to identify the clumpy structures. To perform the image segmentation we utilize {\tt Photutils} package \citep{Bradley20} and define a threshold to choose contiguous pixels ($\geq4$ pixels) that are $\sim 1.5 \sigma$ above the background-subtracted image. The result of this procedure is an image of the clumps of the galaxy. To avoid the pixels that are in the noise level of the galaxy image, we only consider bright, robust clumps by requiring their signal-to-noise to be greater than 3 ($\rm S/N \geq 3$). 

We also note that in the process of clump detection, the central bulge of the galaxy and possible contamination from other bright sources can be falsely detected as clumps. To avoid this, we exclude these objects by requiring a clump to be between $0''.1$ (to remove bulge) and $1.5\times$SMA from the center of the galaxy (to remove nearby sources) defined in F160W band.

\begin{figure*}
\centering
\includegraphics[width=\textwidth,clip=True, trim=0cm 0cm 0cm 0cm]{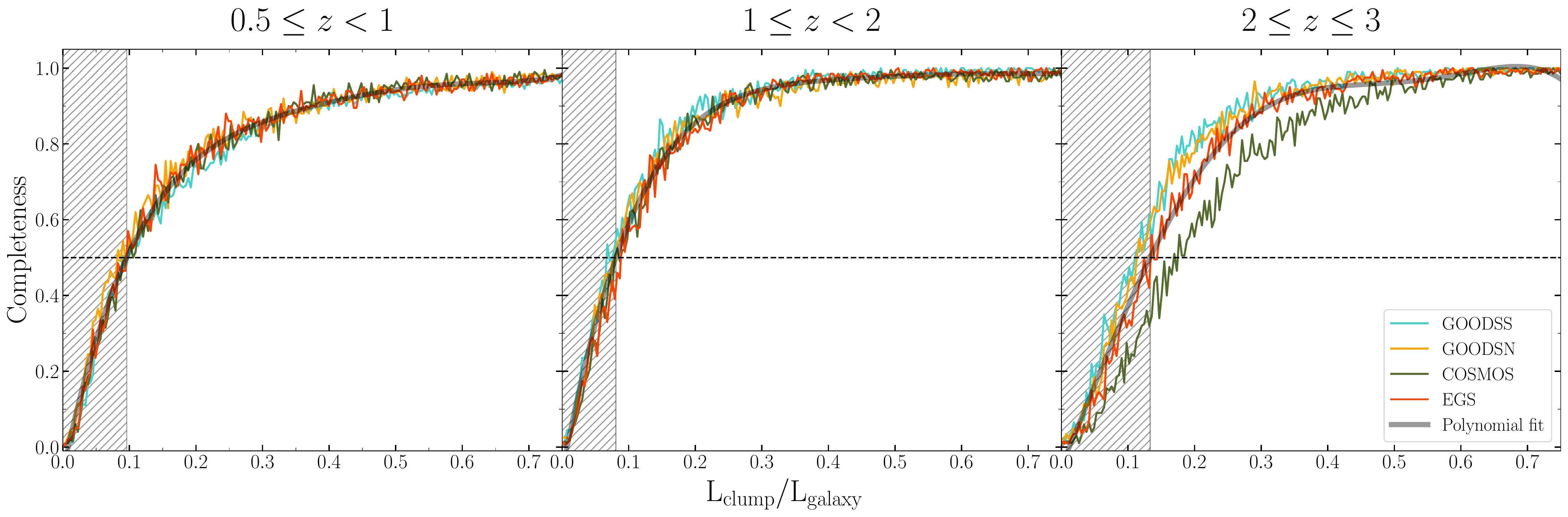}
\caption{To estimate the completeness of our clump identification algorithm, we add one fake clump to each of 800 randomly-selected images of galaxies in four fields at each redshift bin, regardless of being clumpy or non-clumpy, and apply the clump identifier on them to see what percentage of fake clumps are recovered. The luminosity of the pseudo clumps are varied from $0.01\%$ to $75\%$ of the total rest-frame UV luminosity of the host galaxy. The horizontal dashed lines show the success rate of $50\%$ in recovering fake clumps, which corresponds to the clumps that contribute at least $\sim 8-13\%$ to the rest-frame UV light of their host galaxy.}\label{completeness} 
\end{figure*}

Figure \ref{example} shows eight examples of clumpy galaxies with detected clumps after implementing the clump identification method (magenta circles in each image). This figure also shows the F160W images of these clumpy galaxies in the left panels. As we can see, most of these clumps in rest-frame UV images disappeared in the H-band filter. An H-band image primarily reveals the stellar mass distribution of a galaxy, while a rest-frame UV image traces its dust-unobscured SFR.  

\subsection{Success Rate of Clump Identification Method}
\label{complete}

To uniformly define which clumpy galaxies can be detected at all redshifts, we need to determine down to what flux limit we can detect clumps. In order to assess the success rate of our clump identifier, we add one fake clump to the image of the galaxy (whether it is clumpy or not) and test if our clump-finding algorithm recovers the pseudo clump successfully.

Following \cite{Guo15}, we add the pseudo clump to the rest-frame UV image of the galaxy in the desired band using the PSF image of the F160W (H band) band as a point source. The reason to use the H band PSF as a fake clump is that all galaxy images are PSF-matched to this band and thus, the resolution of the rest-frame UV images are similar to that of the H band.

The position of the artificial clump is selected randomly within the size of the galaxy. We vary the flux of the fake clump from $0.01\%$ to $75\%$ of the total rest-frame UV flux of the host galaxy. To test the completeness of the clump finder, we randomly select 200 galaxies in each of the four fields at each redshift bin and perform the process of adding fake clumps. We then feed these galaxy images with fake clumps to our clump identification algorithm to estimate how successful we are in recovering the artificial clumps. Figure \ref{completeness} shows the success rate of clump identifier as a function of the ratio of the clump luminosity to the host galaxy for four fields of GOODS-S, GOODS-N, COSMOS, and EGS in three different redshift bins defined in this work. The horizontal dashed lines show the success rate of $50\%$ which corresponds to the clumps that include at least $\sim 8-13\%$ of the galaxy’s rest-frame UV flux.

Based on the simulations of clump recovery, we define a galaxy as clumpy if it has at least one off-center clump that is brighter than 10\% of the rest-frame UV flux of the galaxy. We conduct experiments with varying threshold values from 8\% to 15\% and find that our results in this study remain consistent, regardless of the chosen 10\% threshold for the clump-to-galaxy flux ratio. Furthermore, we explore using variable thresholds at each redshift to maintain a 50\% completeness rate, yet still observe no significant impact on our findings. In the following section, we perform aperture photometry on the clumps of each galaxy and calculate their rest-frame UV fluxes.

\subsection{Aperture Photometry of Clumps}\label{aper}

We perform fixed aperture photometry on each clump using {\tt Photutils} package \citep{Bradley20}. The radius of the aperture is fixed to be 3 pixels ($0''.18$) for each clump. To estimate the background of an individual clump, we consider two annuli with radii of 6 and 10 pixels around the clump and calculate the mean value of the pixels in between the two annuli. Since there might be contamination from other clumps in the background of each one, we mask all other clumps in a galaxy image while measuring the average background of the clump. We then subtract the average background from the total flux within the aperture radius. Using the PSF in F160W, we calculate that only $\sim 56\%$ of the light from a point source  is encompassed in a 3-pixel aperture radius. Thus, to measure the total flux of the clump, we scale the background-subtracted aperture by this value. 

Using the total flux of individual clumps in the rest-frame UV band, we select those clumps that independently contribute $\geq 10\%$ to the rest-frame UV luminosity of their host galaxy. This value is set based on Figure \ref{completeness} that shows our clump identification method recovers $\sim 50\%$ of the fake clumps with a flux ratio of $\geq 10\%$ compared to the total flux of the host galaxy. The following section presents our results.
\section{Results}\label{Results}

In this section, we investigate the fraction of clumpy galaxies ($\rm f_{clumpy}$) as a function of redshift and also the physical properties of galaxies, such as stellar mass and environment. In order for a galaxy to be considered clumpy, it must have $\geq 1$ off-center clump in its rest-frame UV 1600 \r{A} image. The clumps of galaxies are identified using the methodology described in Section \ref{Clump_identification}. Furthermore, we set a limit on the relative flux of clumps to their host galaxy such that the samples of clumps are $\geq 50\%$ complete when identifying them (Section \ref{complete}). We calculate the fraction of clumpy galaxies, which is the number of clumpy galaxies divided by the total number of galaxies in a given $\rm M_*$, redshift or environment bin. The uncertainty in measuring $\rm f_{clumpy}$ is also calculated using Poisson statistics from the galaxy number count.

\subsection{Redshift Evolution}\label{redshift_result}

\begin{figure}
\centering
\includegraphics[width=0.48\textwidth,clip=True, trim=0cm 0cm 0cm 0cm]{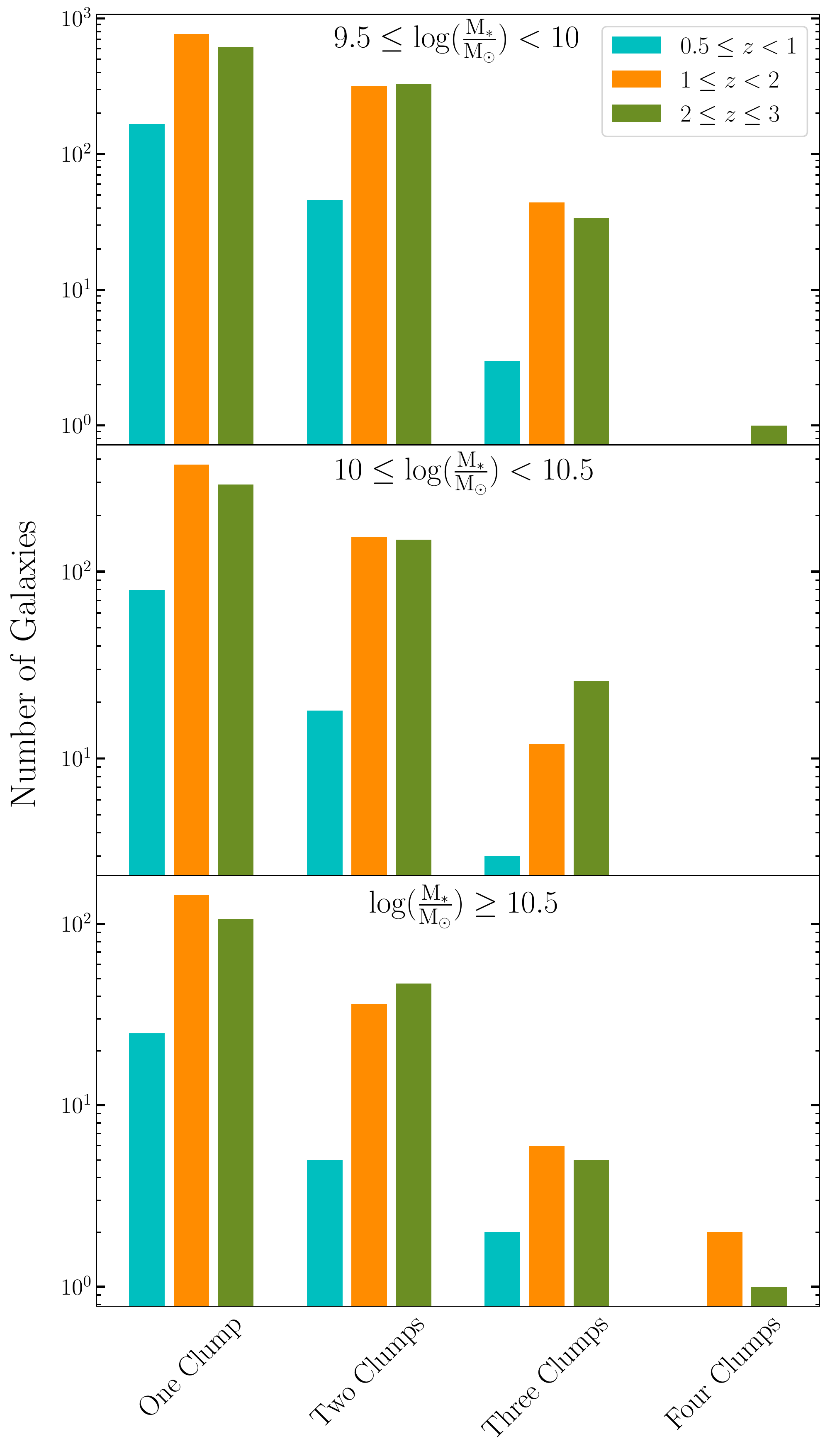}
\caption{From top to bottom, each panel demonstrates the distribution of galaxies with one, two, three and four clumps that contribute $>10\%$ to the rest-frame UV light of their host galaxies in the bins of redshift for three stellar mass ranges $9.5 \leq \rm log(\frac{M_*}{M_\odot})<10$, $10 \leq \rm log(\frac{M_*}{M_\odot})<10.5$ and $\rm log(\frac{M_*}{M_\odot})\geq 10.5$, respectively.}\label{n_clump} 
\end{figure}

Figure \ref{n_clump} illustrates the clump distribution of galaxies in the bins of redshift and stellar mass for all the SFGs selected based on criteria mentioned in Section \ref{sample_selection}, in four fields of GOODS-S, GOODS-N, COSMOS, and EGS. We find that over the redshift range of our study $0.5 \leq z \leq 3$, clumpy galaxies have at most four off-center clumps that contribute more than $10\%$ to the host galaxy's SFR. Moreover, galaxies with a higher number of clumps are mostly found at high redshifts ($z\gtrsim 1$).

\begin{figure}
\centering
\includegraphics[width=0.48\textwidth,clip=True, trim=0cm 0cm 0cm 0cm]{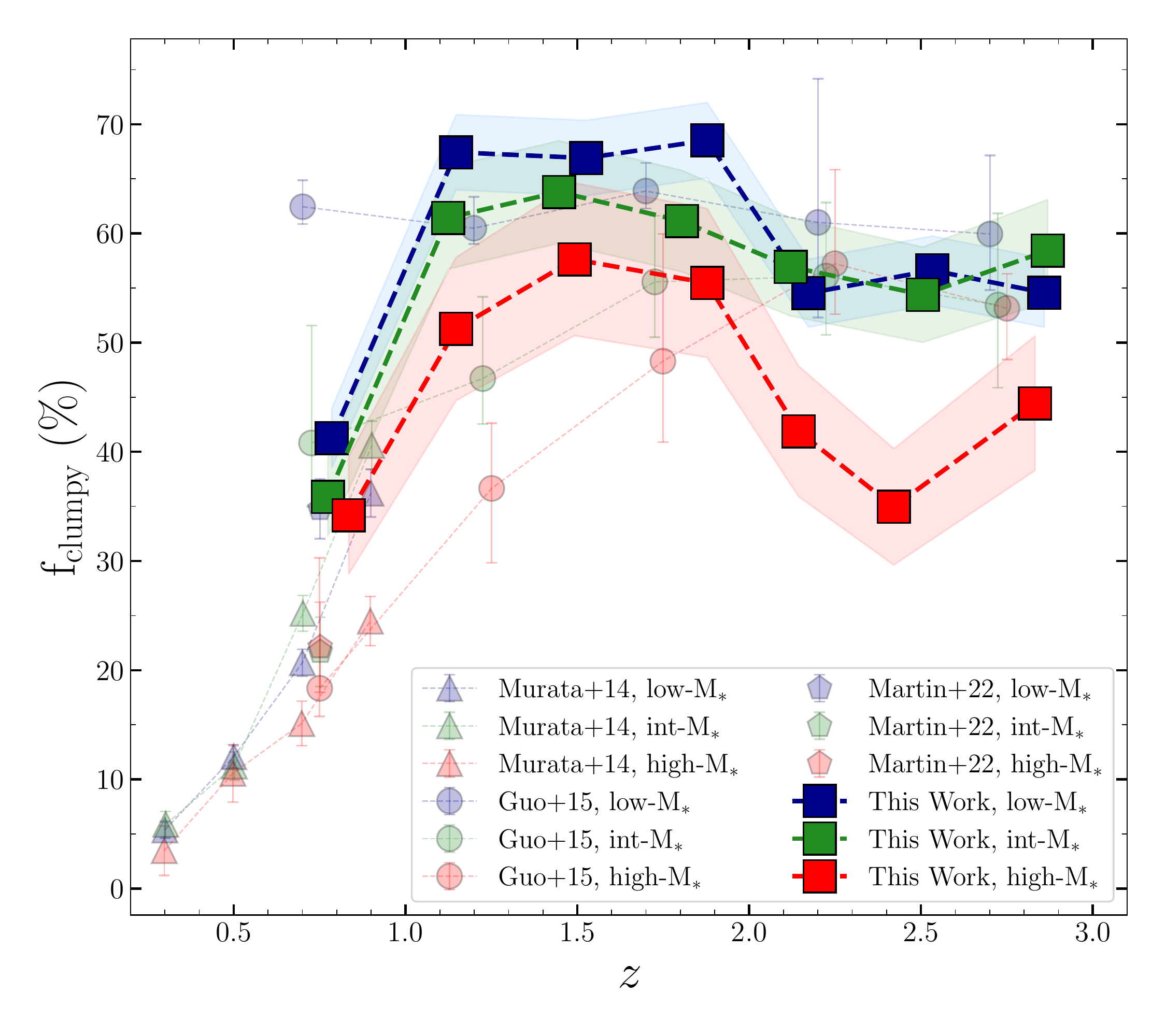}
\caption{Fraction of clumpy galaxies as a function of redshift in three stellar mass bins (squares). Clumpy galaxies are those that have at least one off-center clump in their rest-frame UV images. Shaded regions correspond to $1 \sigma$ uncertainty estimated from Poisson statistics. For comparison, measurements from \cite{Murata14} (triangles), \cite{Guo15} (circles) and Martin et al. (in preparation) (pentagons) are also added. The stellar mass bins in this works are the same as those of \cite{Murata14} and Martin et al. (in preparation) (low-$\rm M_*$: $\rm 9.5 \leq log(\frac{M_*}{M_\odot}) <10$, int-$\rm M_*$: $\rm 10 \leq log(\frac{M_*}{M_\odot}) <10.5$, and high-$\rm M_*$: $\rm log(\frac{M_*}{M_\odot}) \geq 10.5$). But \cite{Guo15} binned the stellar mass of galaxies slightly different (low-$\rm M_*$: $\rm 9 \leq log(\frac{M_*}{M_\odot}) <9.8$, int-$\rm M_*$: $\rm 9.8 \leq log(\frac{M_*}{M_\odot}) <10.6$, and high-$\rm M_*$: $\rm 10.6 \leq log(\frac{M_*}{M_\odot}) <11.4$).}\label{z_evol} 
\end{figure}

Figure \ref{z_evol} shows the redshift evolution of $\rm f_{clumpy}$ in three bins of stellar mass. The bins of redshift are considered such that in each bin, we have the same number of galaxies. As seen in Figure \ref{z_evol}, the fraction of clumpy galaxies as a function of redshift follows almost the same trend for all three stellar mass bins of low-mass, intermediate-mass and high-mass galaxies (blue, green, and red squares for galaxies with the stellar mass of $\rm 9.5 \leq log(\frac{M_*}{M_\odot}) <10$, $\rm 10 \leq log(\frac{M_*}{M_\odot}) <10.5$, and $\rm log(\frac{M_*}{M_\odot}) \geq 10.5$, respectively), with $\rm f_{clumpy} \sim 35\%-60 \%$ at $z \sim 2.5$ increasing to $\sim 55\%-70 \%$ at $z \sim 1.5$, and decreasing to $\sim 35\%-40 \%$ at lower redshifts ($z <1$). Furthermore, galaxies in higher stellar mass bin tend to have lower clumpy fraction compared to those with intermediate and lower stellar masses, especially at $z \sim 2.5$, where this fraction is $\sim 25 \%$ lower for massive galaxies.

\begin{figure*}
\centering  
\includegraphics[width=0.66\linewidth]{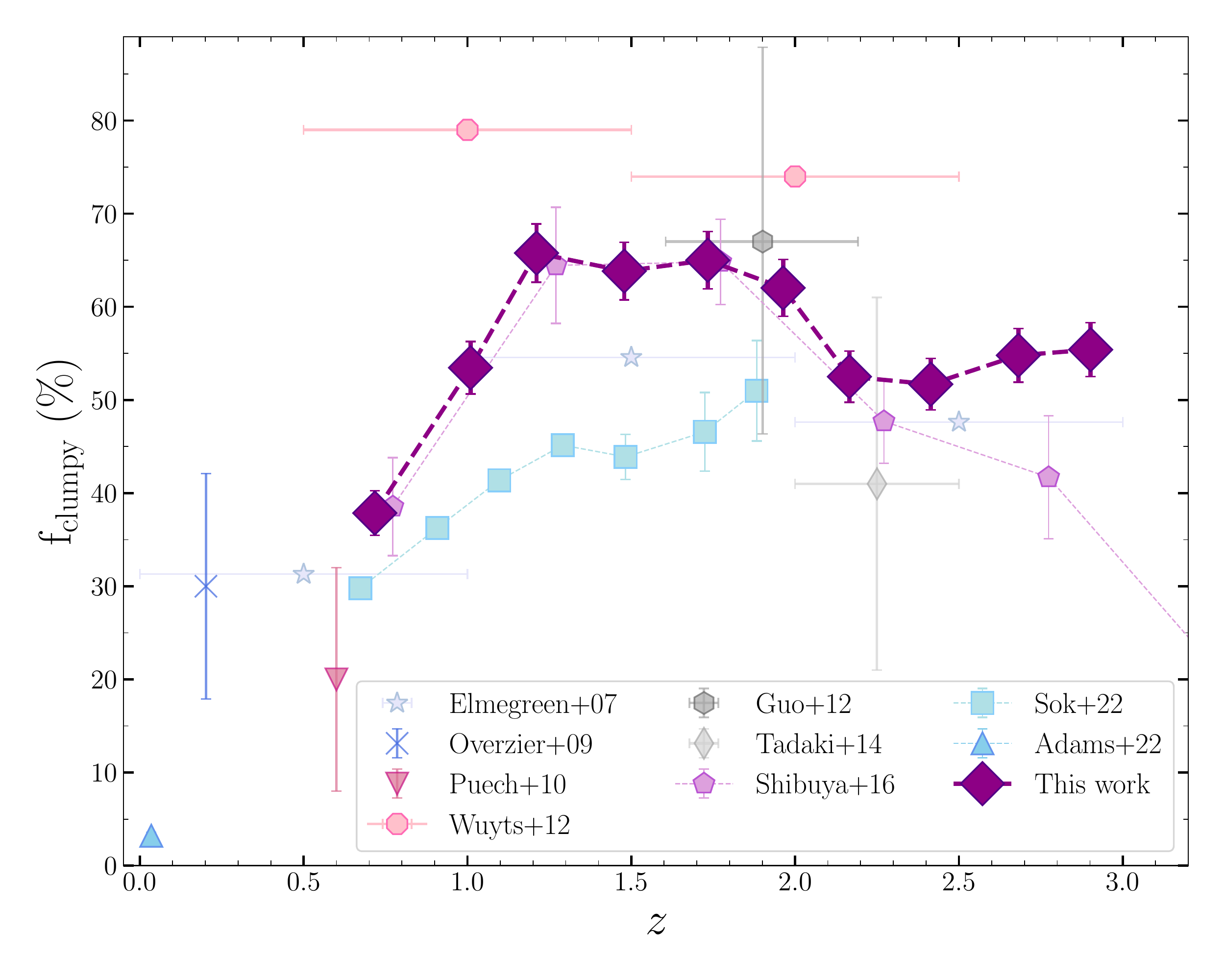}
\caption{Same as Figure \ref{z_evol}, but the fraction of clumpy galaxies is not binned by stellar masses. Also, more studies of clumpy fraction are added to the figure for comparison. Summary of previous studies on clumpy galaxies is presented in Table \ref{tab:prev}.}
\label{z_evol_comp}
\end{figure*}

Additionally, we present $\rm f_{clumpy}$ as a function of redshift for the entire sample regardless of their stellar masses in Figure \ref{z_evol_comp}. We find that the fraction of clumpy galaxies (shown with purple diamonds) is at its highest ($\sim 65\%$) in the redshift range $\sim 1-2$ and decreases to $\sim 40\%$ towards lower redshifts. Also, the trend is almost constant beyond $z\gtrsim 2$ with $\rm f_{clumpy} \sim 50\%$.

As can be seen in Figure \ref{z_evol} — and repeated in Figure \ref{z_evol_comp} with the average for the whole sample — there is a steep decline in the clumpiness of SFGs at late times, in a manner that appears to reflect the decline in star formation rate density (SFRD) \citep[e.g.,][]{Madau14}. At early times, the data do not simply reflect the early rise in SFRD and there may be a number of issues that complicate the measurement and interpretations of clumpiness at those epochs.

To compare our clumpy fraction measurements with previous works, we show measurements of $\rm f_{clumpy}$ in different redshifts from other studies in Figure \ref{z_evol} and \ref{z_evol_comp} and summarize some of them in Table \ref{tab:prev}. Nevertheless, clumps are defined differently in different studies, and various wavelengths can be used to identify them, so the comparison between different studies is qualitative.

\cite{Murata14} studied the evolution of $\rm f_{clumpy}$ as a function of redshift and reported an increase of clumpy galaxies with increasing redshift for galaxies at $0.2<z<1$. Our result is in general agreement with them in the low- and intermediate-mass bins. However, in the high-mass bin, we find higher fraction of clumpy galaxies. Part of the discrepancy between our measurements and their values can be due to the detection band of clumps and the definition of clumpy galaxies. While \cite{Murata14} considered at least three clumps as the criteria to select a clumpy galaxy in optical (HST/ACS F814W) images of galaxies, we identify clumps in the rest-frame UV maps and determine clumpy galaxies with at least one off-center clump.

Another comprehensive study on the fraction of clumpy galaxies is done by \cite{Guo15} with a sample of SFGs at $0.5<z<3$ in the two CANDELS/GOODS-S and UDS fields. They utilized rest-frame UV 2500 \r{A} HST/ACS images of galaxies and detected clumps through an algorithm that identifies high-intensity pixels. A threshold of $8\%$ is employed on the luminosity of the individual clumps relative to the host galaxy. Moreover, \cite{Shibuya16} took a sample of SFGs, and Lyman break galaxies (LBGs) in five CANDELS fields, Hubble Ultra Deep Field and eXtreme Deep Field \citep[HUDF and XDF;][]{Beckwith06,Bouwens11,Illingworth13,Ellis13,Koekemoer13} and the parallel fields of Abell 2744 and MACS0416 in the Hubble Frontier Fields \citep[HFF;][]{Coe15,Atek15,Oesch15,Ishigaki15} with a similar sample selection to \cite{Guo15} and investigated the fraction of clumpy galaxies at $0<z<8$. They used the same definition of clumps and the detection algorithm as \cite{Guo15}. 

\begin{table*}[t]
\centering

\caption{Summary of previous studies on clumpy galaxies in field.}

\begin{tabular*}{\textwidth}{@{\extracolsep{\fill}}cccccc@{}}

\hline
Reference       & Sample ($\rm N_{galaxy}$)                       & $\rm log(\frac{M_*}{M_\odot})$ \footnote{Stellar mass range of sample used in each study.} & Redshift  & Method & Detection Band                           \\ \hline \hline
\cite{Elmegreen07}       & Starbursts (1003)                            & N/A                 & 0-5       & Visual                         & F775W                                    \\
\cite{Overzier09}        & LBAs (20)                                    & 9-10                & $\sim$0.2 & Visual                         & $\rm UV_{rest}$                             \\
\cite{Puech10}           & Emission-line galaxies (63)                  & $>10$    & $\sim$0.6 & Visual                         & F435W                                    \\
\cite{Guo12}             & SFGs (10)                                    & $>10$    & 1.5-2.5   & Algorithm                      & F850LP                                   \\
\cite{Wuyts12}           & SFGs (649)                                   & $>10$    & 0.5-2.5   & Algorithm                      & rest-frame 2800 \r{A}                        \\
\cite{Tadaki14}          & HAEs (100)                                   & 9-11.5              & 2-2.5     & Algorithm                      & F606W \& F160W                           \\
\cite{Murata14}          & $\rm I_{F814W}<22.5$ galaxies (24027) & $>$9.5   & 0.2-1     & Algorithm                      & F814W                                    \\
\cite{Guo15}             & SFGs (3239)                                  & 9-11.5              & 0.5-3     & Algorithm                      & $\rm UV_{rest}$ 2500 \r{A}                      \\
\cite{Shibuya16}         & Photo-z galaxies \& LBGs (16910)             & 9-12                & 0-8       & Algorithm                      & $\rm UV_{rest}$ \& $\rm opt_{rest}$                                 \\
\cite{Huertas20} & SFGs (1500)                                  & 9-11.5              & 1-3       & Algorithm                      & $\rm UV_{rest}$ \& $\rm opt_{rest}$                                        \\
\cite{Adams22}             & SFGs (58550)                                 & $>9$   & 0.02-0.15     & Visual                      & $ugriz$                                        \\
\cite{Sok22}             & SFGs (20185)                                 & $>9.8$   & 0.5-2     & Algorithm                      & $\rm U_{rest}$, $\rm V_{rest}$ \& $\rm opt_{rest}$                                        \\
Martin et al. (in preparation)          & SFGs (695)                                   & $>9.5$   & 0.5-1     & Algorithm                      & $\rm UV_{rest}$ 1600 \r{A}                      \\
This work       & SFGs (6767)                                 & $>9.5$   & 0.5-3     & Algorithm                      & $\rm UV_{rest}$ 1600 \r{A} \\ \hline
\end{tabular*}%
\label{tab:prev}
\end{table*}

Although our mass bins are slightly different from the ones employed by \cite{Guo15}, our result on the evolution of $\rm f_{clumpy}$ for intermediate-mass galaxies is in qualitative agreement with theirs (green circles in Figure \ref{z_evol}), which shows an increase of $\rm f_{clumpy}$ with increase in redshift and then flattening around $z \sim 1.5$. However, this fraction in their low-mass bin is almost independent of redshift at all redshift bins (blue circles). In contrast, our result for this mass bin follows the same trend as our intermediate-mass bin. In the high-mass bin, \cite{Guo15} reported a monotonically increasing $\rm f_{clumpy}$ with redshift (red circles), while our measurement agrees on the increase out to $z \sim 1.5$ and then shows a sign of decrease at higher redshifts. We speculate that part of this discrepancy, at least in the low-mass bin at low redshifts, is due to the fact that \cite{Guo15} detected clumps in rest-frame Near-UV (NUV) images, while our study is conducted on the Far-UV (FUV) images. Recently, a study by Martin et al. (in preparation) is using UVCANDELS data to study demographics of clumpy galaxies at $0.5\leq z \leq 1$ with a similar clump identification method to \cite{Guo15} on the FUV images of galaxies. Their measurements of $\rm f_{clumpy}$ are demonstrated by pentagons in Figure \ref{z_evol}. At the redshift range of their study, our low-mass bin measurements are in agreement with theirs within the uncertainties.

Moreover, reported measurements of $\rm f_{clumpy}$ as a function redshift by \cite{Shibuya16} are shown in Figure \ref{z_evol_comp} with pentagons. Our estimates agree well with theirs out to $z\sim 2$. However, with increasing redshift, their fraction of clumpy galaxies decreases while ours flattens. Other studies of clumpy galaxies have also investigated the evolution of clumpy galaxy fraction with redshift. For instance, a recent study by \cite{Sok22} detected clumpy galaxies at $0.5<z<2$ by deconvolving ground-based images of galaxies in the COSMOS field to increase their resolution, and measured the fraction of clumpy galaxies (blue squares in Figure \ref{z_evol_comp}). Similar to ours, their $\rm f_{clumpy}$ is increasing with redshift below $z \sim 2$.

In conclusion, the decline in clumpiness from cosmic noon to today is very well measured, regardless of technique. It is, therefore, possible that part of the decline in cosmic SFRD can be attributed to the decline in the prevalence of clumps, which are sites of star formation. The formation and evolution of clumps can be affected by various physical mechanisms, including both internal processes (e.g., stellar feedback, AGN activity) and/or external processes (e.g., galaxy interactions, strangulation). To constrain the dominant process responsible for clump formation and evolution, in the following sections, we study $\rm f_{clumpy}$ as a function of stellar mass and environment. Feedback processes scale with stellar mass of galaxies, therefore, any relationship between stellar mass and clumpy fraction indicates that internal processes are responsible for clump evolution, while any correlation with the environment indicates that external processes play a role.

\subsection{Stellar Mass Dependence}\label{mass_result}

\begin{figure}
    \centering
	\includegraphics[width=0.48\textwidth,clip=True, trim=0cm 0cm 0cm 0cm]{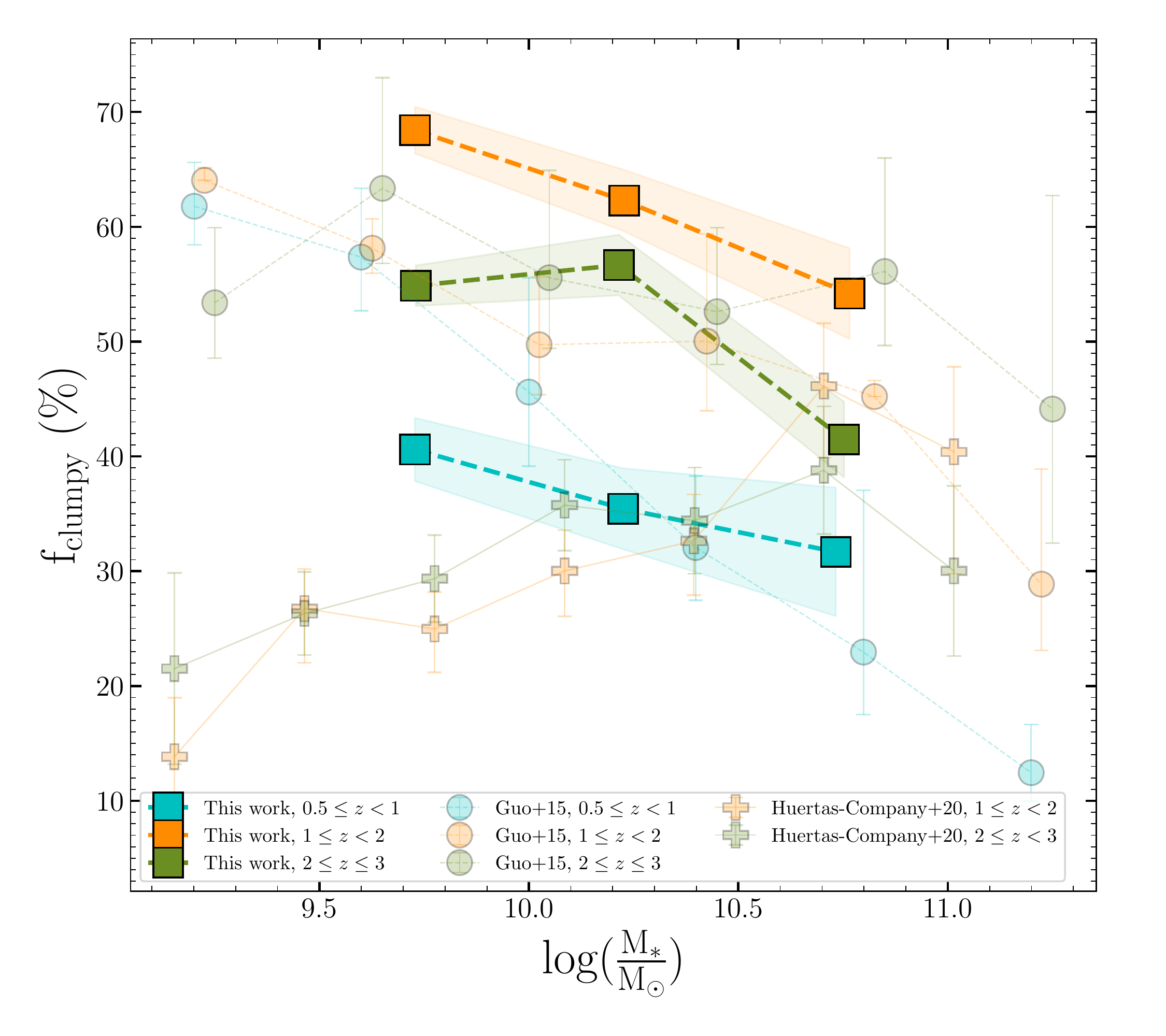}
	\caption{Fraction of clumpy galaxies as a function of stellar mass in different redshift bins. $1 \sigma$ uncertainties of $\rm f_{clumpy}$ measurements are shown with shaded regions that are calculated by Poisson statistics. Circle and plus data points show reported measurements from \cite{Guo15} and \cite{Huertas20}, respectively.}\label{m_evol} 
\end{figure}

We present the fraction of clumpy galaxies in redshift bins as a function of stellar mass in Figure \ref{m_evol}. At all three redshift bins of $0.5 \leq z <1$ (cyan squares), $1 \leq z <2$ (orange squares), and $2 \leq z \leq 3$ (green squares) the fraction of clumpy galaxies decreases monotonically with increase in stellar mass. At low redshifts, $\rm f_{\rm {clumpy}}$ decreases from $\sim 40\%$ for galaxies with stellar masses in the range $9.5 \leq \rm log(\frac{M_*}{M_\odot})<10$ to $\sim 30\%$ for massive galaxies ($\rm log(\frac{M_*}{M_\odot}) \geq 10.5$). The slope of this decrease is steeper for galaxies at $z>1$. At $1 \leq z<2$, the fraction of clumpy galaxies decreases from $\sim 70\%$ to $\sim 55\%$ for galaxies with $9.5 \leq \rm log(\frac{M_*}{M_\odot})<10$ and $\rm log(\frac{M_*}{M_\odot}) \geq 10.5$, respectively. At the highest redshift bin, $\rm f_{clumpy}$ is almost independent of stellar mass ($\sim 55\%$) for low-mass galaxies, but drops quickly to $\sim 40\%$ in massive galaxies.

\begin{figure*}
    \centering
    \subfloat{{\includegraphics[width=0.47\textwidth,clip=True, trim=0cm 0cm 0cm 0cm]{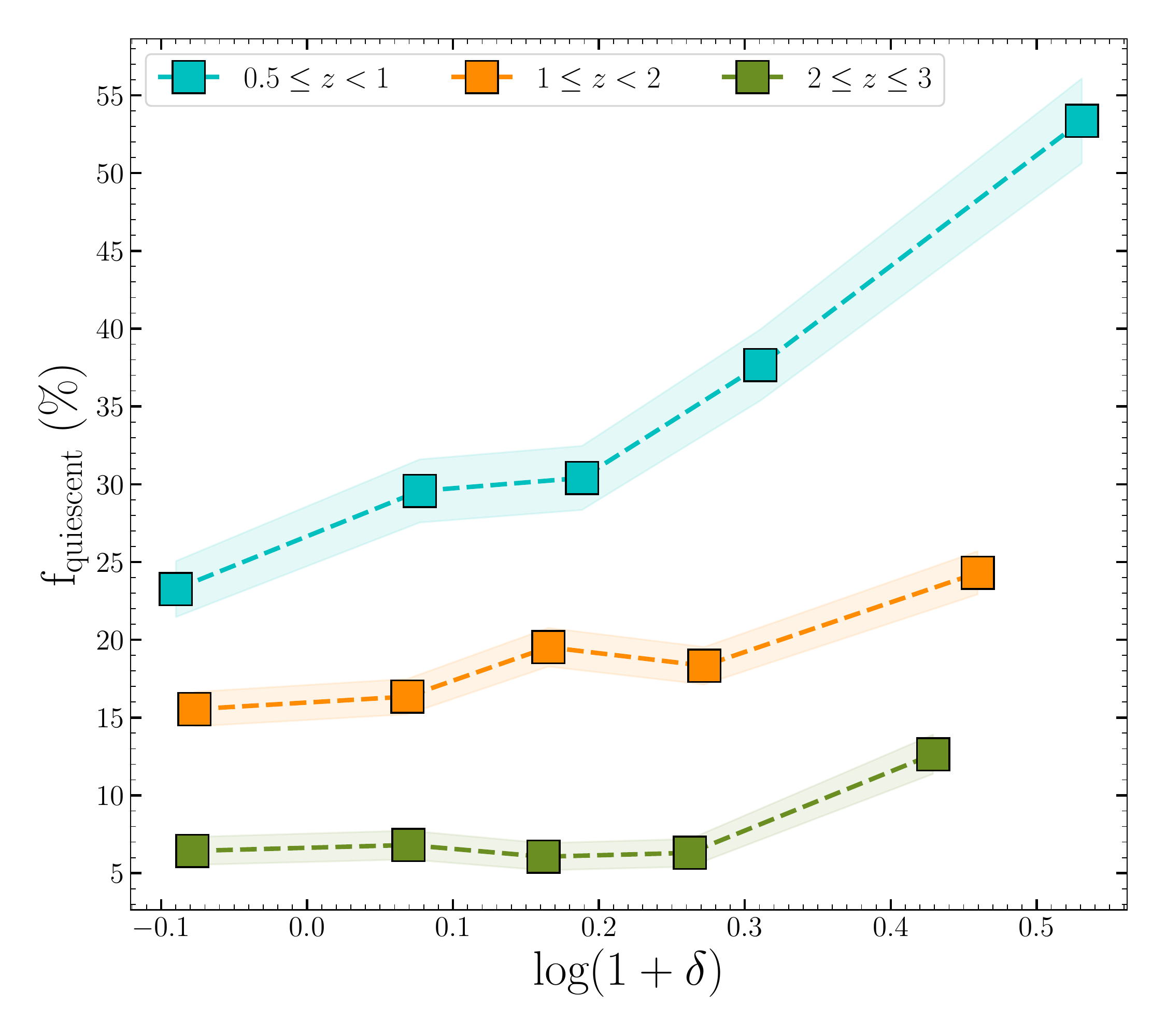} }}%
    \subfloat{{\includegraphics[width=0.47\textwidth,clip=True, trim=0cm 0cm 0cm 0cm]{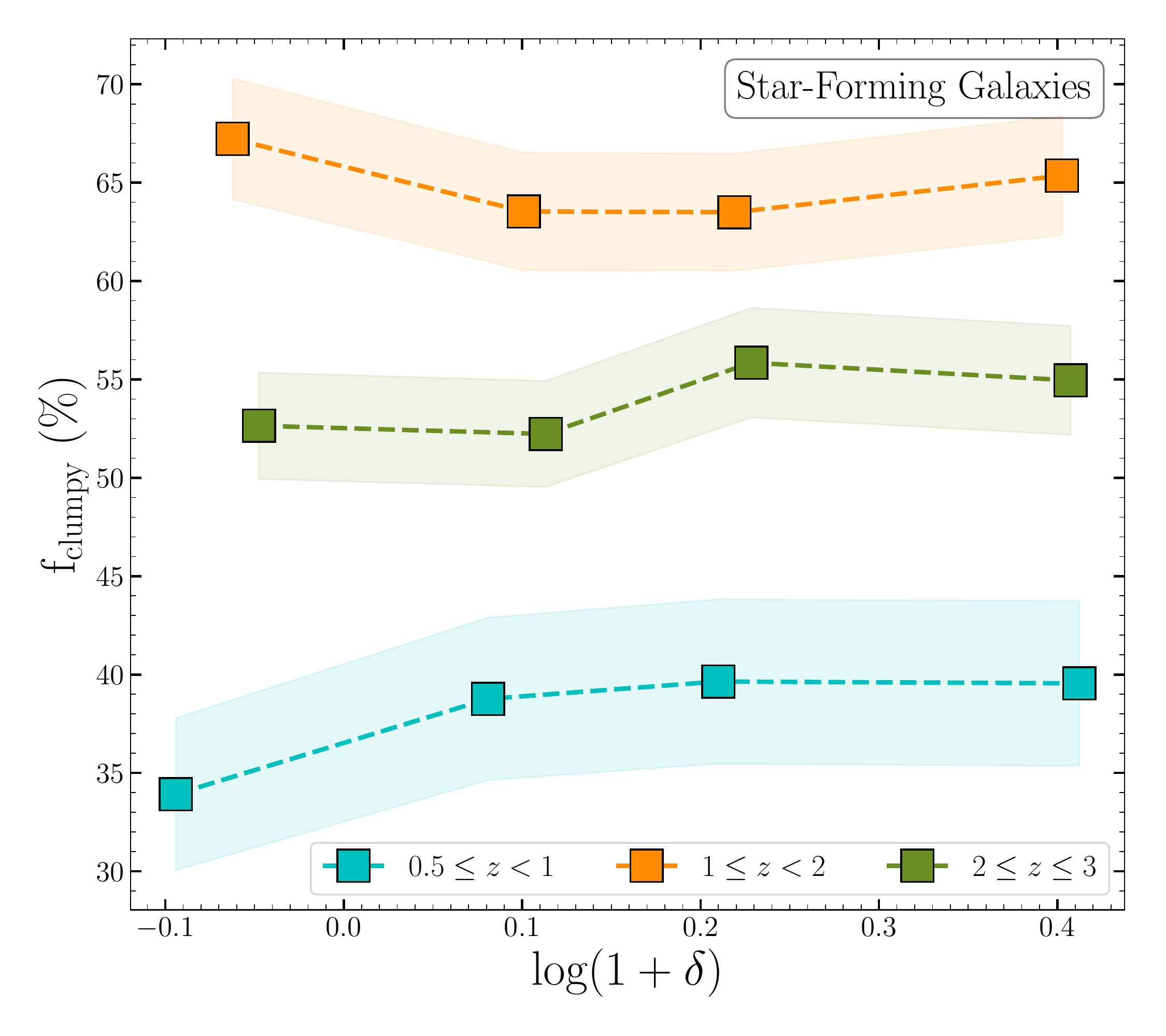} }}%
    \qquad
    \caption{\textit{Left}: Fraction of quiescent galaxies in different bins of redshift and environment. \textit{Right}: Fraction of clumpy galaxies in SFGs as a function of environment in different redshift bins. Error bars (shaded regions) for both panels are estimated by Poisson statistics from the number count of galaxies.}\label{env_evol} 
\end{figure*}

For comparison, two other studies which have demonstrated the fraction of clumpy galaxies as a function of stellar mass are shown in Figure \ref{m_evol} as well. \cite{Guo15} found that $\rm f_{clumpy}$ is decreasing with the increase of stellar mass (cyan, orange and green circles), which is in agreement with our results. Moreover, \cite{Huertas20} used neural networks to detect clumps in the rest-frame optical and UV images of galaxies in the CANDELS fields and hydrodynamic zoom-in simulations of VELA \citep{Ceverino14} at $1<z<3$. They measured the fraction of clumpy galaxies as a function of stellar mass and redshift and found that $\sim 40\%$ of galaxies with $\rm log(\frac{M_*}{M_\odot})>10$ are clumpy, and this fraction drops to $\sim 20\%$ for low-mass ($\rm log(\frac{M_*}{M_\odot})<10$) galaxies (orange and green pluses). This result is in contrast with our clumpy fraction and those reported in \cite{Guo15}.

\subsection{Environmental Dependence}\label{env_result}

It has been shown, at least at low redshifts ($z \lesssim 1$), that star formation activity of galaxies is strongly correlated with their surrounding environments \citep{Patel09,Peng10,Darvish16,Chartab20}. Thus, clumps, which are sites of star formation in galaxies, may be linked to the local environment of their host galaxies, and by studying the environmental dependence of $\rm f_{clumpy}$, one can gain insights into clump formation and evolution. 

Several simulations and observational studies suggest that clumps join the disk of galaxies through minor mergers, and their formation is \textit{ex-situ} \citep{Hopkins13,Straughn15}. As minor/major mergers are more prevalent in dense environments \citep{Hine16,Watson19}, examining the correlation between clumps and their host galaxies' local environment would constrain the clump formation mechanisms. It is possible that \textit{ex-situ} formation is the dominant process of clump formation in galaxies if the clumpy fraction is correlated with the local environment.

\begin{figure*}
    \centering
	\includegraphics[width=\textwidth,clip=True, trim=3cm 3cm 4cm 3cm]{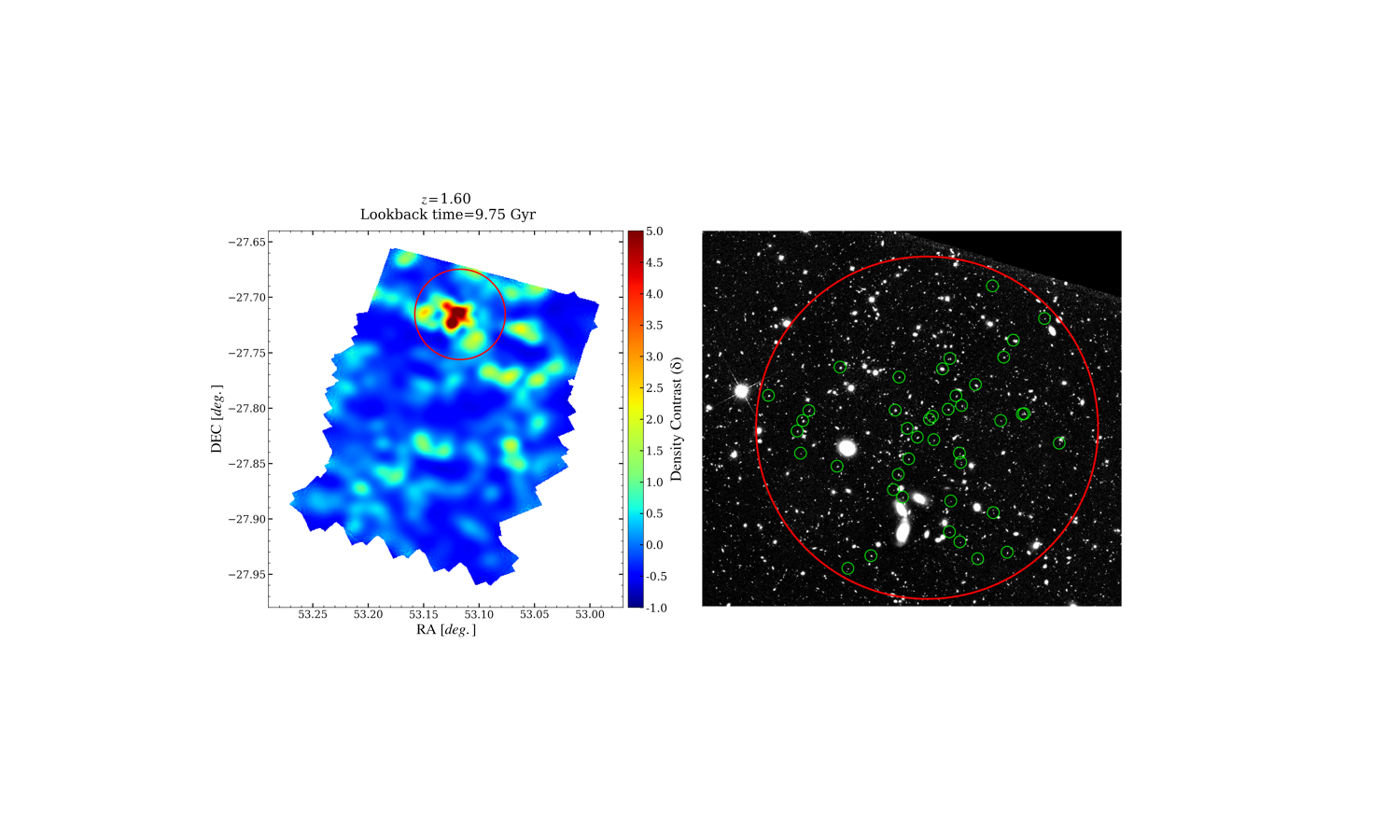}
	\caption{\textit{Left}: Density map of the GOODS-S field at the redshift of the spectroscopically-confirmed cluster adopted from \cite{Chartab20}, which is color-coded by density contrast. The density map confirms the existence of this over-density at $z=1.61$ in the red clustered region. The red circle with the radius of $\rm 1\ Mpc$ (physical)  indicates the boundary of this cluster which encloses the cluster members. \textit{Right}: The footprint of this cluster at $z=1.61$ in the sky. Green circles show the members of the cluster, while the red one is the same as the red circle in the left panel.}\label{env_cluster} 
\end{figure*}

Our sample of galaxies, located in four CANDELS fields, has local environmental measurements available from the catalog of \cite{Chartab20}, which includes measurements of local density for galaxies with $\rm HST/F160W \leq 26$ AB mag in all the CANDELS fields. They reconstructed density maps of galaxies probabilistically using the weighted kernel density estimation method in a wide redshift range ($0.4<z<5$). Their density measurements are based on uniformly calculated photometric redshifts with well-calibrated probability distributions across the CANDELS fields \citep{Kodra22}. In this work, we utilize the local density catalog of \cite{Chartab20} to study the fraction of clumpy galaxies in different environments. For a full description of environment measurements, we refer readers to \cite{Chartab20,chartab21}. We note that due to the limited size of CANDELS fields, these density measurements only probe the galaxy groups and cores of the structures rather than extended structures such as protoclusters \citep{chartab21}.

It is well-known that the fraction of quiescent galaxies is positively correlated with the local density contrast ($\delta$), especially at $z\lesssim1$ \citep{Patel09,Peng10,Balogh16,Kawin17,Darvish17,Chartab20}. The left panel in Figure \ref{env_evol} demonstrates the fraction of quiescent galaxies as a function of local environment ($1+\delta$) for a sample of galaxies with $\rm F160W \leq 25$ mag and $\rm M_*\geq 10^{9.5}M_\odot$ in three redshift bins used in the present work. We define quiescent galaxies with a cut on sSFR as described in Section \ref{sample_selection}. Across all redshift bins, we find a positive correlation, which is stronger at the lowest redshift bin ($0.5\leq z <1$). We also estimate the fraction of clumpy galaxies in three redshift bins as a function of their local environment for a sample of SFGs described in Section \ref{sample_selection}. The right panel of Figure \ref{env_evol} shows that the fraction of clumpy galaxies is almost independent of the environment in all three redshift bins of $0.5 \leq z<1$ (cyan squares), $1 \leq z<2$ (orange squares), and $2 \leq z \leq 3$ (green squares). $\rm f_{clumpy}$ is constant around $\sim 35\%-40\%$, $\sim 50\%-55\%$, and $\sim 65\%$ from lower to higher redshift bin, respectively. It suggests that in dense environments, there are rapid processes that quench galaxies before interfering with their clumps.

Moreover, the lack of a significant relationship between the clumpy fraction and the environment of galaxies may indicate that clumps are rather formed \textit{in-situ} than \textit{ex-situ}. It is possible, however, that measurements of clumpy fractions or galaxies' environments suffer large uncertainties, and the correlation, if any, is too weak to be detected by these measurements.

To assess this issue, we study the fraction of clumpy galaxies within a spectroscopically-confirmed cluster in the GOODS-S field around the redshift where the fraction of clumpy galaxies reaches its maximum ($z \sim 1.5$). This cluster has 42 spectroscopically-confirmed members at $z_{\rm med}=1.61$ within $\Delta z=0.01$, which is a virialized structure with X-ray detection \citep{Kurk09}. Figure \ref{env_cluster} shows the footprint of this cluster and its confirmed members in the sky (right panel), as well as the density map of GOODS-S field at $z=1.6$ reconstructed by environment measurements of \cite{Chartab20} color-coded by the density contrast (right panel). The red region in the left panel shows this over-density with $>4\sigma$ significance. On the right panel, confirmed members of the cluster are identified by green circles, while the red circle indicates the boundary of the cluster, which has a radius of $\rm 1\ Mpc$ (physical).

\begin{figure}
    \centering
	\includegraphics[width=0.48\textwidth,clip=True, trim=0cm 0cm 0cm 0cm]{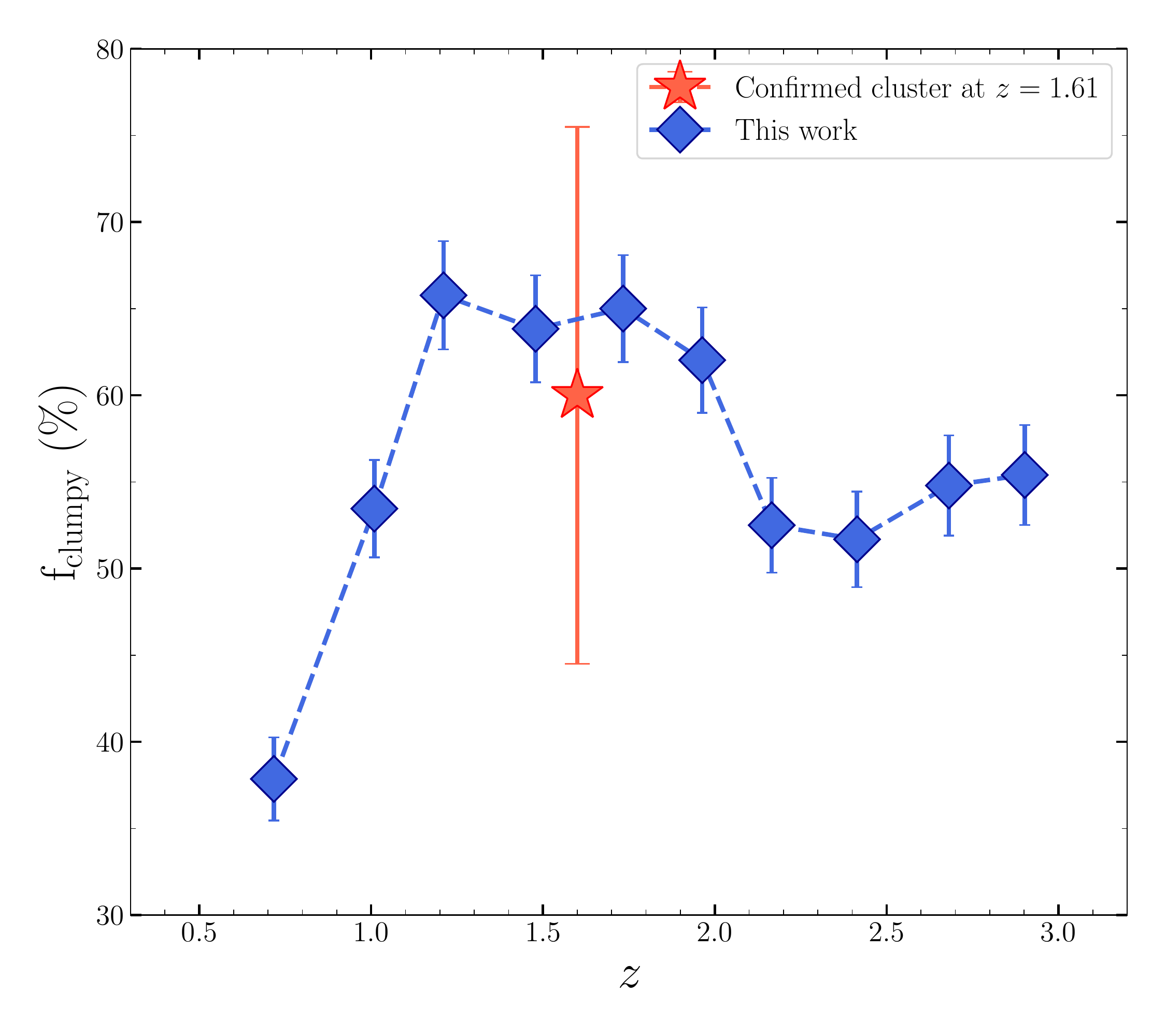}
	\caption{Same as Figure \ref{z_evol_comp}, but the fraction of clumpy galaxies for a confirmed cluster at $z=1.61$ is also shown with a red star. The error bars are measured using Poisson statistics from the number counts of galaxies.}\label{z_evol_cluster} 
\end{figure}

Out of 42 members of this cluster, 25 were in the stellar mass range of our study ($\rm log(\frac{M_*}{M_\odot}) \geq 9.5$). We perform clump detection analysis on the cluster members and found that 15 out of 25 of them are classified as clumpy, corresponding to an $\rm f_{clumpy}$ of $60\%$ for this structure. We overlaid the measurement of $\rm f_{clumpy}$ for this cluster in the figure showing the fraction of clumpy galaxies as a function of redshift for our entire sample (Figure \ref{z_evol_cluster}). We find less than $5\%$ discrepancy in the clumpy fraction of this structure compared to our total sample, which is insignificant and within the measurement uncertainties. 

Our assessment of the environmental measurements by studying a spectroscopically-confirmed cluster confirm the lack of trend in Figure \ref{env_evol}, implying that $\rm f_{clumpy}$ is independent of the environment of galaxies. However, further studies are needed to confirm this for a statistically large sample of structures. The availability of future wide surveys with deep and high-resolution images will facilitate such studies as they will enable reliable measurements of the environment and deep-resolved images of galaxies.

\section{Summary}\label{Discussion}

In this paper, we identify star-forming clumpy sub-structures in the rest-frame UV 1600 \r{A} images of SFGs selected from four CANDELS fields. The rest-frame UV at the redshift range of our study ($0.5 \leq z \leq 3$) is probed by F275W, F435W and F606W filters for which the observations are conducted by Hubble Space Telescope via UVCANDELS and CANDELS surveys. We utilize low-pass band filter in Fourier space to reconstruct the background image of the galaxies and subtract them from the galaxy images to detect clumps. We study the fraction of clumpy galaxies (the number of galaxies with at least one detected off-center clump in their images divided by the total number of galaxies) as a function of their host galaxies' physical properties, such as stellar mass and local environment. We also investigate the clumpy fraction evolution with redshift and compare our results with that of previous works. Moreover, clump statistics of a spectroscopically-confirmed cluster at $z=1.61$ is calculated for 25 members. Our findings can be summarized as follows:

\begin{itemize}
    \item We find that the fraction of clumpy galaxies peaks at redshifts $\sim 1-2$ ($\rm f_{clumpy} \sim 65\%$), and decreases to $\sim 40\%$ at lower redshifts. Furthermore, the fraction of clumpy galaxies is almost redshift independent beyond $z\gtrsim 2$ with $\rm f_{clumpy} \sim 50\%$.  

    \item The fraction of clumpy galaxies decreases monotonically with an increase in stellar mass. The slope of this decline is steeper for galaxies at higher redshifts ($z>1$).

    \item For the first time, we study the fraction of clumpy galaxies as a function of their local environment derived from accurate photometric redshifts out to $z=3$. We find that $\rm f_{clumpy}$ is independent of local environment of galaxies across the redshift range of this study ($0.5 \leq z \leq 3$). We also investigate the clumpy fraction for the members of a spectroscopically-confirmed cluster at $z=1.61$. Out of 25 selected members of this cluster in the stellar mass range of our study, 15 are labeled as clumpy, resulting in the $\rm f_{clumpy}=60\%$ for this cluster. This result is consistent with our measurements of clumpy fraction in the field at the same redshift.

    \item Due to the lack of a significant correlation between the clumpy fraction and the local environment of galaxies, it appears that clump formation is facilitated by the fragmentation of gas clouds under VDI rather than being caused by incidents in the local environment of galaxies (e.g., mergers).

\end{itemize}


The analysis in this paper relies on observations with the NASA/ESA Hubble Space Telescope obtained at the Space Telescope Science Institute, which is operated by the Association of Universities for Research in Astronomy, Incorporated, under NASA contract NAS5- 26555. Support for Program number HST-GO-15647 was provided through a grant from the STScI under NASA contract NAS5-26555.

\bibliography{clump}

\end{document}